\documentclass[]{aa}
\usepackage{graphicx}
\begin{document}

\title{The effect of ram pressure on the star formation, mass distribution and morphology of galaxies}

\titlerunning{How does ram pressure affect galaxies?}

\author{W. Kapferer\inst{1}
   \and C. Sluka\inst{1}
   \and S. Schindler\inst{1}
   \and C. Ferrari\inst{1,2}
   \and B. Ziegler\inst{3}}

\offprints{W. Kapferer, \email{wolfgang.e.kapferer@uibk.ac.at}}

\institute{1 Institut f\"ur Astro- und Teilchenphysik,
            Universit\"at Innsbruck, Technikerstrasse 25,
            A-6020 Innsbruck\\
            2 Laboratoire Cassiop\'ee, CNRS, UMR 6202, Observatoire de la C\^ote d Azur,
            BP4229, 06304 Nice Cedex 4\\
            3 European Southern Observatory, Karl-Schwarzschild-Strasse 2, D-85748 Garching}

\date{}

\abstract {}{We investigate the dependence of star formation and the
distribution of the components of galaxies on the strength of ram
pressure. Several mock observations in X-ray, H$\alpha$ and HI
wavelength for different ram-pressure scenarios are presented.}{By
applying a combined N-body/hydrodynamic description (GADGET-2) with
radiative cooling and a recipe for star formation and stellar
feedback 12 different ram-pressure stripping scenarios for disc
galaxies were calculated. Special emphasis was put on the gas within
the disc and in the surroundings. All gas particles within the
computational domain having the same mass resolution. The relative
velocity was varied from 100 km/s to 1000 km/s in different
surrounding gas densities in the range from $1\times10^{-28}$ to
$5\times10^{-27}$ g/cm$^3$. The temperature of the surrounding gas
was initially $1\times10^{7}$ K.}{The star formation of a galaxy is
enhanced by more than a magnitude in the simulation with a high
ram-pressure ($5\times10^{-11}$ dyn/cm$^2$) in comparison to the
same system evolving in isolation. The enhancement of the star
formation depends more on the surrounding gas density than on the
relative velocity. Up to 95\% of all newly formed stars can be found
in the wake of the galaxy out to distances of more than 350 kpc
behind the stellar disc. Continuously stars fall back to the old
stellar disc, building up a bulge-like structure. Young stars can be
found throughout the stripped wake with surface densities locally
comparable to values in the inner stellar disc. Ram-pressure
stripping can shift the location of star formation from the disc
into the wake on very short timescales. As the gas in a galaxy has a
complex velocity pattern due to the rotation and spiral arms, the
superposition of the internal velocity field and the ram pressure
causes complex structures in the gaseous wake which survive
dynamically up to several 100 Myr. Finally we provide simulated
X-ray, H$\alpha$ and HI observations to be able to compare our
results with observations in these wavebands. These simulated
observations show many features which depend strongly both on the
strength and the duration of the external ram pressure.}{}


\maketitle

\section{Introduction}

Since the late 1970's observations have revealed an evolution of
galaxy cluster members with redshift. It has been found that the
fraction of star-forming and post-star-forming systems increases
significantly with redshift (Dressler et al. 1987, 1999). In
addition Butcher \& Oemler (1978) showed that there is a strong
evolution from bluer to redder colours in cluster galaxies. They
have found an excess of blue objects at z = 0.5 with respect to
lower redshift systems, the so-called Butcher$-$Oemler effect. Many
different mechanisms have been studied with special emphasis on
their effect on the morphology and the star formation of cluster
galaxies. Among these mechanisms ram-pressure stripping (e.g. Gunn
\& Gott 1972), galaxy harassment (e.g. Moore et al. 1998) or
strangulation/starvation (e.g. Larson et al. 1980) can be found. How
strong these different processes influence the evolution of cluster
galaxies is still under debate. It is crucial to identify and
disentangle different interaction processes, also at intermediate
and high redshift. One possible way is to study the evolution of the
star formation activity in galaxies and the dependence of this
quantity on the galaxies environment (Balogh et al. 2002, Gerken et
al. 2004, Poggianti et al. 2006, Verdugo et al. 2008). As the
physical processes involved in star formation are poorly understood,
additional properties of galaxies were investigated to study whether
a galaxy has an interaction with its environment or not. The total
gravitational potential of galaxies and its distortions due to
interactions with their direct surroundings can be used to
disentangle between external and internal processes of galaxy
evolution. Technically it is nowadays possible to observe the full
2D velocity field (VF) of local galaxies in optical wavebands using
integral field units (IFUs), i.e. SAURON (e.g. Ganda et al. 2006) or
Fabry-Perot interferometry
(e.g. Chemin et al. 2006; Garrido et al. 2002).\\
Another approach are numerical simulations to study individually the
effects of the various processes. Kronberger et al. (2006, 2007)
investigated in numerical simulations the effects of galaxy-galaxy
mergers and tidal interactions between galaxies on the internal
kinematics of galaxies. The results of these simulations were also
used to identify observational biases in observations of the
velocity field of distant galaxies (Kapferer et al. 2006; Kronberger
et al. 2007). Although dependencies on the viewing angle and on the
spatial resolution have been found, it was shown that tidal
interactions mainly introduce nonaxisymmetric and nonbisymmetric
features (see also Rubin et al. 1999).\\
From combined X-ray and optical observations of galaxy clusters it
was found that the so called intra-cluster medium (ICM), a hot
($10^{7}<$T$<10^{8}$ K) thin ($\rho\sim1\times10^{-27}$ g/cm$^{3}$)
plasma, harbours five times more mass than the galaxies themselves.
This plasma exerts a pressure on the inter-stellar medium (ISM) of
the cluster galaxies, which removes gas from the disc if the force
due to the external ram pressure exceeds the restoring gravitational
force of the galaxy. The effect of ram-pressure stripping on
individual galaxies was investigated by several groups. The
dependence of the stripping radius on the external ram pressure and
the galaxy properties was investigated (e.g. Abadi et al. 1999;
Vollmer et al. 2001; Roediger \& Hensler 2005) and a remarkable
agreement between analytical estimates and numerical simulations was
found. Further galaxy-scale numerical simulations for ram-pressure
stripping were presented by Quilis et al. (2000), Mori \& Burkert
(2000), Toniazzo \& Schindler (2001), Schulz \& Struck (2001), and
Roediger \& Br\"uggen (2006, 2007). Jachym et al. (2007) used an
N-body/SPH code to investigate the influence of a time varying ram
pressure on spiral galaxies. A closer investigation of the stripped
gaseous wake with Eulerian grid techniques were carried out by
Roediger \& Br{\"u}ggen (2008). Recently also the effects of the
external ram pressure on the star formation rate of the stripped
galaxy were studied (Kapferer et al. 2008; Kronberger et al. 2008).
The influence of the stripped gas from the galaxies affected by ram
pressure on the metallicity of the ICM was investigated by Schindler
et al. (2005), Domainko et al. (2006) and Kapferer et al. (2007) in
large scale combined
N-body/hydrodyanmic simulations.\\
In this work we investigate the dependence of the star formation in
a galaxy on different strengths of ram pressure. We study the
influence of different strengths of ram pressure on the distribution
of the components of a galaxy, i.e. the resulting stellar and gas
distribution. In addition we provide mock observations in the radio,
the optical and the X-ray range.

\section{The simulation setup}

\subsection{The hydrodynamic description}
The simulations were carried out with the N-body/SPH code GADGET-2
developed by V. Springel (see Springel 2005 for details). The code
treats the gas of the galaxies and the surrounding gas by smoothed
particle hydrodynamics (SPH Gingold \& Monaghan 1977; Lucy 1977).
The collisionless dynamics of the dark matter and the stellar
component is modelled by an N-body technique. Additional routines
for cooling, star formation (SF), stellar feedback, and galactic
winds are included as described in Springel \& Hernquist (2003).

\subsection{The initial model for the model galaxies}
\begin{figure}
\begin{center}
\includegraphics[width=5cm]{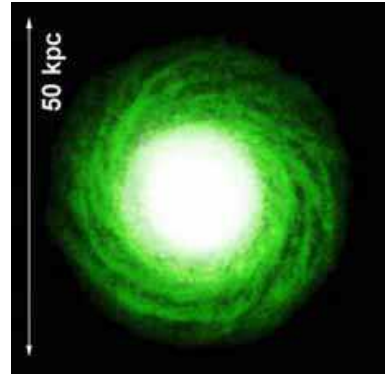}
\caption{The distribution of gaseous (green) and stellar (white)
mass of the model galaxy after 2 Gyr of isolated evolution.}
\label{gas_stars}
\end{center}
\end{figure}

The model galaxies were created with an initial disc galaxy
generator developed by Volker Springel. Details and analysis can be
found in Springel et al. (2004). The model galaxy has a stellar
disc-scale length of 3.3 kpc and a halo circular velocity of 160
km/s. The initial gas fraction in the disc is 25\% of the total disc
mass. The mass resolution of the different components of the
galaxies (gas, stellar, dark matter) is listed in Table
\ref{model_galaxy}. In Fig. \ref{gas_stars} the gas and stellar
distribution of the model galaxy is shown after 2 Gyr of evolution
without external ram pressure.

\begin{table}
\caption[]{Initial properties of the model galaxy}
\begin{tabular}{c | c c c c c}
\hline \hline & number of & mass resolution & total mass\cr  &
particles & [M$_{\odot}$/particle] & [M$_{\odot}$] \cr \hline DM
halo & $3\times10^5$& $3.5\times10^6$ & $1.05\times10^{12}$\cr
gaseous disc & $2\times10^5$ & $3.4\times10^4$ & $6.80\times10^9$\cr
stellar disc & $2\times10^5$ & $1.0\times10^5$ & $2.00\times10^{10}$\cr
\hline
\end{tabular}
\label{model_galaxy}
\end{table}

\begin{figure}
\begin{center}
\includegraphics[width=\columnwidth]{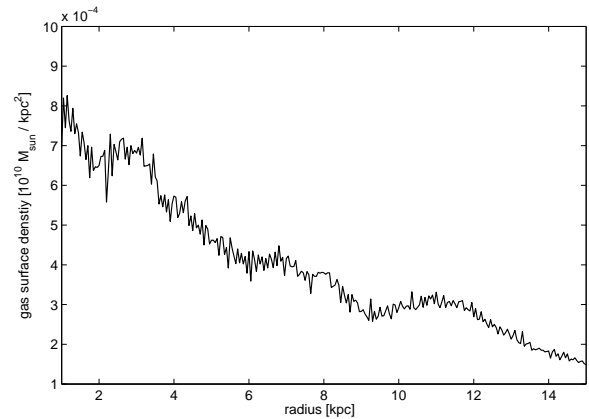}
\caption{The radial gas profile of the model galaxy after 2 Gyr of
isolated evolution.} \label{gas_profile_isolated}
\end{center}
\end{figure}

After 2 Gyr of isolated evolution, i.e. no surrounding gas,
$2.2\times10^{10}$ M$_{\sun}$ new stars have formed and the actual
star formation rate is 1.6 M$_{\sun}$/yr. In Fig.
\ref{gas_profile_isolated} the density profile of the gas for the
model galaxy after 2 Gyr of evolution is shown. The several local
maxima originate from the pattern of the gaseous spiral arms.

\subsection{The star formation and feedback model}
We applied, as in Kronberger et al. (2008) and Kapferer et al.
(2008), the so called hybrid method for star formation and stellar
feedback, introduced by Springel \& Hernquist (2003). The
fundamental assumption of this model is the conversion of cold
clouds into stars on a characteristic timescale $t_*$ and the
release of a certain mass fraction $\beta$ due to supernovae (SNe).
The minimum temperature the gas can reach due to radiative cooling
is $10^{4}$K. From observations and analytical models it was found,
that matter can escape the galaxies potential due to thermal and/or
cosmic-ray driven winds caused by SN explosions (Breitschwerdt et
al. 1991). We applied the same method as  Springel \& Hernquist
(2003) and selected the mass of the outflow such, that the mass of
the galactic wind is direct proportional to the actual star
formation rate with a proportionally factor of two. More details on
the model can be found in Springel \& Hernquist (2003).

\subsection{The ram pressure}
\begin{figure}
\begin{center}
\includegraphics[width=\columnwidth]{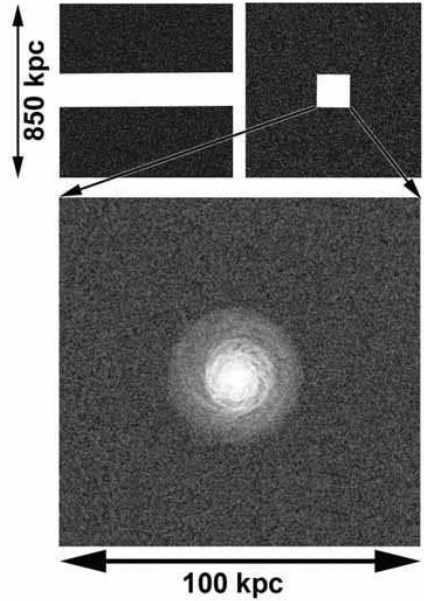}
\caption{The simulation box with the high resolution gaseous wind
tunnel in the middle, surrounded by a low resolution region. In the
upper panels the computational domain is shown from the side and
from the top, whereas in the lower panel a the inner (highly
resolved) wind tunnel with the galaxy is seen. The greyscale
reflects the particle density.} \label{simulation box}
\end{center}
\end{figure}

\begin{figure}
\begin{center}
\includegraphics[width=\columnwidth]{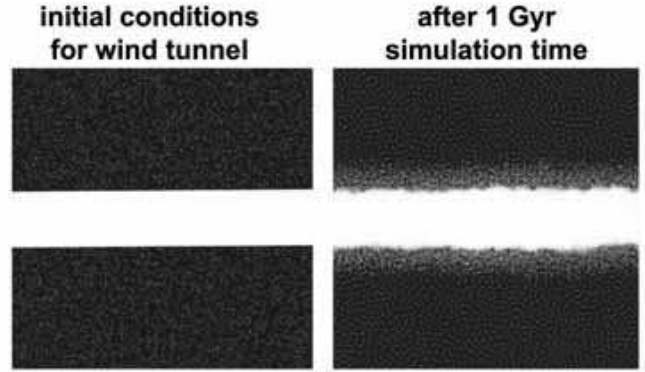}
\caption{The simulation box with the high resolution gaseous wind
tunnel in the middle shown at the start of the simulation and after
1 Gyr of evolution. The greyscale reflects the particle density.}
\label{simulation box_evo}
\end{center}
\end{figure}

To study the dependence of the star-formation rate and the
morphology of a galaxy on the strength of an external ram pressure,
we expose the model galaxy after 2 Gyr of isolated evolution to an
external wind. The galaxy moves then with a constant velocity
through the ambient medium in a face-on orientation. The simulations
are set up in the same way as in Kronberger et al. (2008) and
Kapferer et al. (2008) with a modification regarding the mass
resolution of the gas particles. The simulation box has 850 kpc on a
side and periodic boundaries. In order to obtain the same mass
resolution for the gas in the model galaxy and in the surrounding
gas a high resolution wind tunnel is set in the simulation box,
which is surrounded by a less resolved medium, in order to keep the
wind tunnel's hydrodynamic properties stable during the simulation
time which is 1 Gyr. In Fig. \ref{simulation box} the simulation box
is shown.
In Fig. \ref{simulation box_evo} the gas particle distribution of
the surrounding gas at the beginning and after 1 Gyr of evolution
for simulation 1 (see Table \ref{ram_pressure_table}) is shown. The
mean density of the wind tunnel does not change more than 4\% over 1
Gyr of simulation time for all chosen density configurations, which
leads to a nearly constant ram pressure on the model galaxy during
the whole simulation.\\
Table \ref{ram_pressure_table} gives the details on the strength of
the ram pressure for all simulations. In total, twelve different
strengths of ram pressure acting face-on on the model galaxy are
simulated. The values for the surrounding gas densities and the
relative velocities represent a wide range of ram pressure acting on
galaxies in galaxy clusters. We know from X-ray observations of
galaxy clusters, that the density from the ICM is in the range of
several $10^{-26}$ g/cm$^{-3}$ (i.e. several 10$^{-2}$ cm$^{-3}$ in
number density) in the central regions down to several $10^{-28}$
g/cm$^{-3}$ (i.e. several 10$^{-4}$ cm$^{-3}$ in number density) in
the outskirts (Schindler et al. 1999; Russel et al. 2008), depending
on the total mass of the galaxy cluster. From optical
spectroscopical investigations it is evident that galaxies have
velocities of up to several 1000 km/s in the centre of galaxy
clusters, depending on the dynamical state of the galaxy cluster and
the total mass of the system, e.g. Maurogordato et al. (2008). We
split our simulation into three major groups, first the fast moving
galaxies with a relative velocity of 1000 km/s, second an
intermediate relative velocity of 500 km/s and third a group of very
slow moving galaxies with a relative velocity of 100 km/s with
respect to the ambient medium. Note that the temperature of the
ambient medium is fixed to $1\times10^{7}$ K.
\begin{table}
\caption[]{The properties of the ram pressure and the surrounding
gas of the simulations.}
\begin{tabular}{c | c c c c}
\hline \hline  & v$_{\rm{rel}}$ & $\rho_{\rm{sur}}$ & particle & p
\cr run & [km/s] & [g/cm$^{3}$] & number & [dyn cm$^{-2}$] \cr
\hline 1 & 1000 & $1\times10^{-28}$ & $1.08\times10^{6}$ &
$1.00\times10^{-12}$ \cr 2 & 1000 & $5\times10^{-28}$ &
$5.40\times10^{6}$ & $5.00\times10^{-12}$ \cr 3 & 1000 &
$1\times10^{-27}$ & $1.08\times10^{7}$ & $1.00\times10^{-11}$ \cr 4
& 1000 & $5\times10^{-27}$ & $5.40\times10^{7}$ &
$5.00\times10^{-11}$ \cr 5 & 500 & $1\times10^{-28}$ &
$1.08\times10^{6}$ & $2.50\times10^{-13}$ \cr 6 & 500 &
$5\times10^{-28}$ & $5.40\times10^{6}$ & $1.25\times10^{-12}$ \cr 7
& 500 & $1\times10^{-27}$ & $1.08\times10^{7}$ &
$2.50\times10^{-12}$ \cr 8 & 500 & $5\times10^{-27}$ &
$5.40\times10^{7}$ & $1.25\times10^{-11}$ \cr 9 & 100 &
$1\times10^{-28}$ & $1.08\times10^{6}$ & $1.00\times10^{-14}$ \cr 10
& 100 & $5\times10^{-28}$ & $5.40\times10^{6}$ &
$5.00\times10^{-14}$ \cr 11 & 100 & $1\times10^{-27}$ &
$1.08\times10^{7}$ & $1.00\times10^{-13}$ \cr 12 & 100 &
$5\times10^{-27}$ & $5.40\times10^{7}$ &
$5.00\times10^{-13}$\cr\hline
\end{tabular}
\label{ram_pressure_table}
\newline v$_{\rm{rel}}$...relative velocity between the surrounding
gas and the galaxy; $\rho_{\rm{sur}}$...surrounding gas density;
p...ram pressure on the galaxy.\\
\end{table}
How turbulence, especially Kelvin-Helmholtz instabilities, would
affect the gaseous structures in the wake of the stripped galaxies
is discussed in Kapferer et al. (2008). Concluding from this
previous investigation Kelvin-Helmholtz instabilities affect dense,
cool gas knots in the wake, but the Kelvin-Helmholtz timescales for
the given densities and the relative velocities are much larger than
the dynamical timescales of ram-pressure stripping. This leads to
the conclusion that the structures in the gaseous wake can at least
survive within the simulation time, which is typically 750 Myr.

\section{Results}
In this section we investigate the influence of different strengths
of ram pressure on the gaseous and stellar components of the model
galaxy. In addition we present mock observations of the simulations
in different wavelengths regimes. The virtual observations span from
radio to optical and X-ray wavelengths.

\subsection{The influence of ram pressure on the star formation}
\begin{figure}
\begin{center}
\includegraphics[width=\columnwidth]{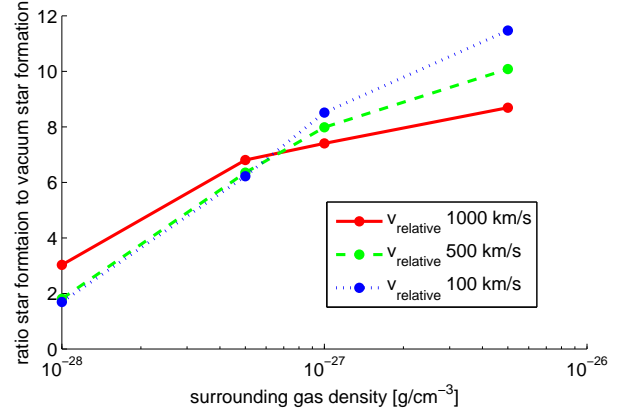}
\caption{Ratio of the star formation of the different simulations to
the star formation integrated over 500 Myr for the isolated evolving
galaxy.} \label{ratio_star_formation_vacuum}
\end{center}
\end{figure}

\begin{figure}
\begin{center}
\includegraphics[width=\columnwidth]{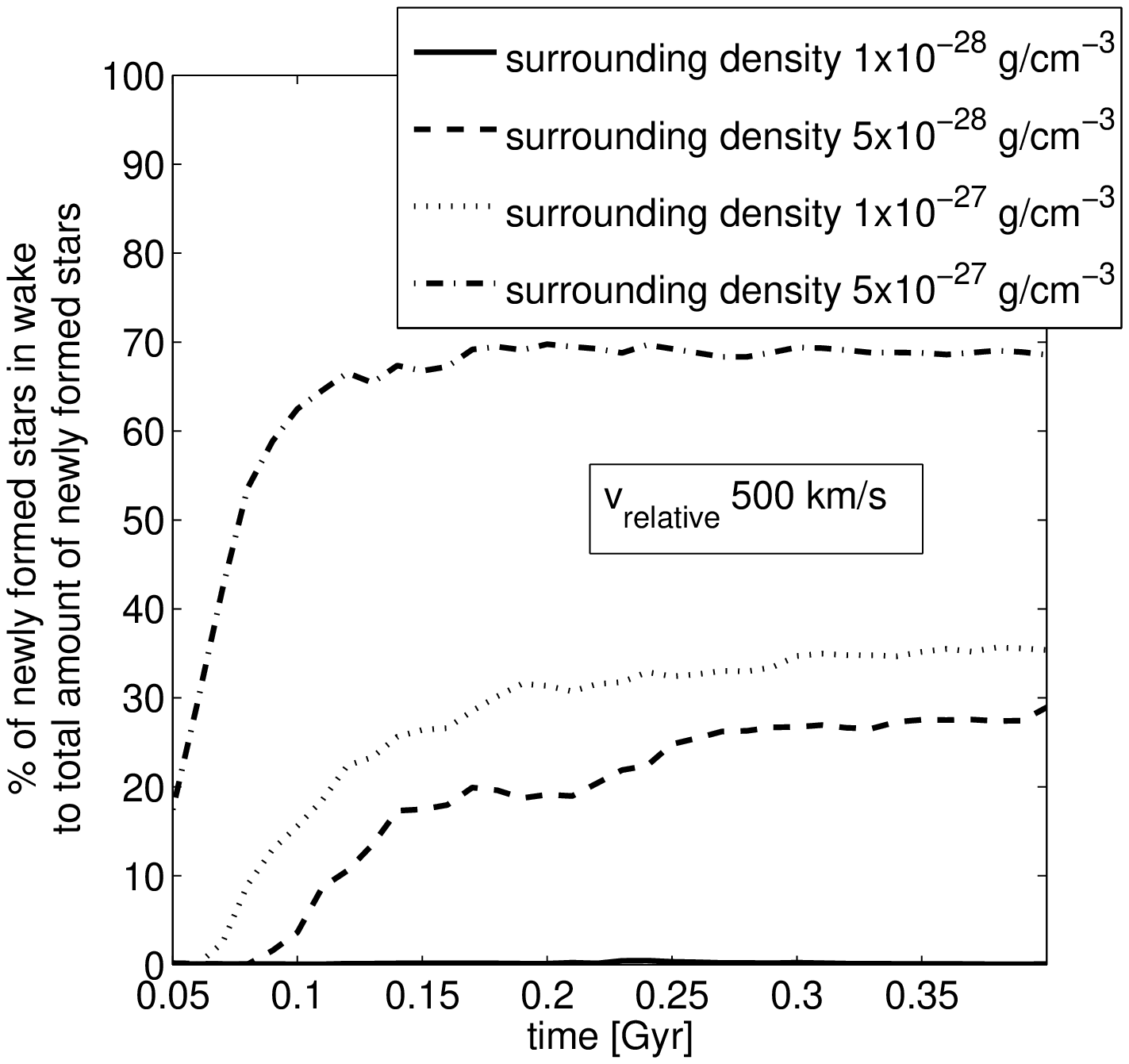}
\caption{The evolution of the mass fraction of newly formed stars in
the wake (distances larger than 10 kpc above the disc) in comparison
to the total amount of newly formed stars for v$_{\rm{rel}}$=500
km/s. Note that the solid line is, corresponding to the surrounding
gas density of $1\times10^{-28}$ g/cm$^{-3}$, is almost constantly
zero, i.e. no stars are formed at distances larger than 10 kpc
behind the disc.}
\label{newly_formed_stars_in_wake_to_total_500km/s}
\end{center}
\end{figure}

\begin{figure}
\begin{center}
\includegraphics[width=\columnwidth]{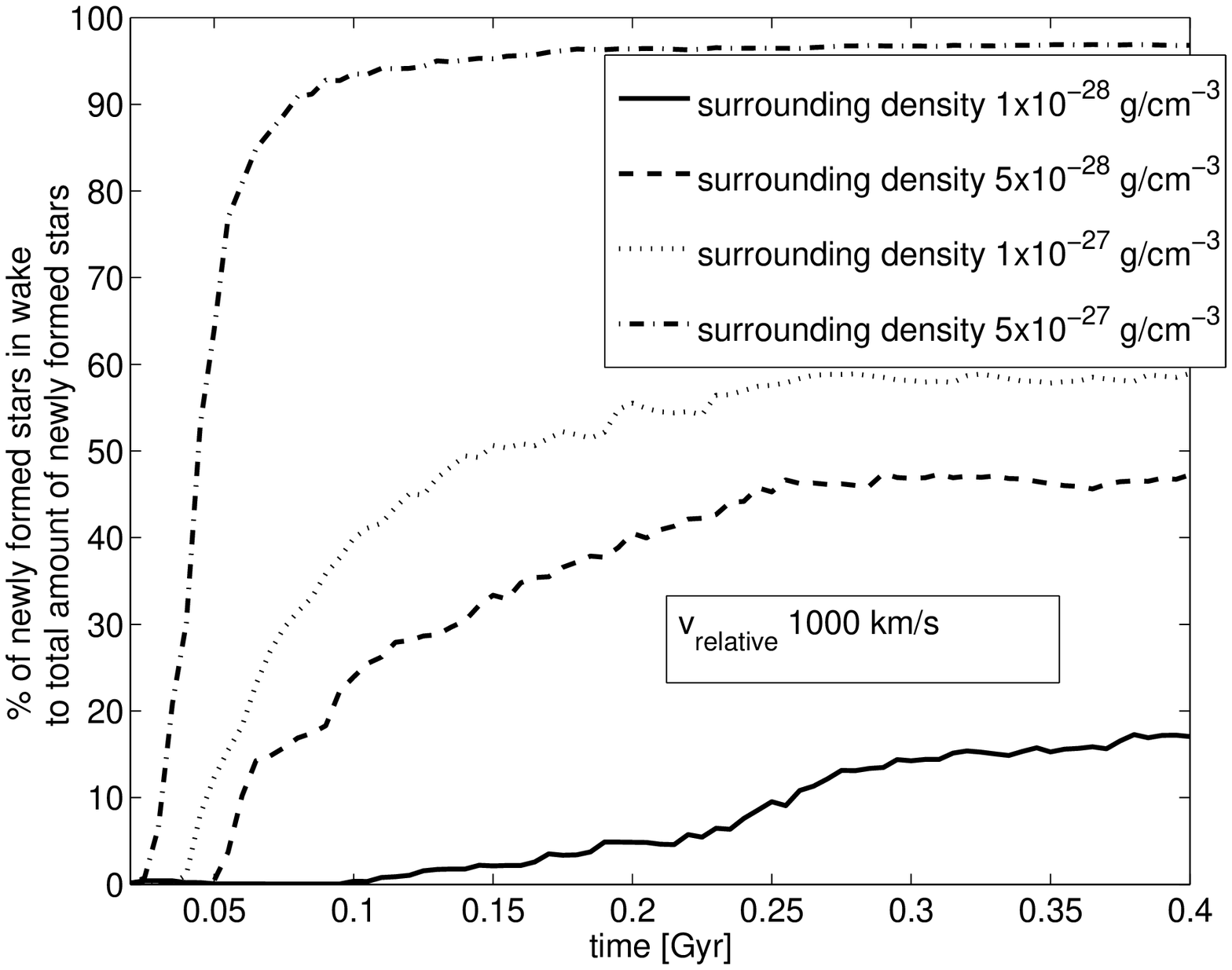}
\caption{The evolution of the fraction of newly formed stars in the
wake (distances larger than 10 kpc above the disc) in comparison to
the total amount of newly formed stars for v$_{\rm{rel}}$=1000
km/s.} \label{newly_formed_stars_in_wake_to_total_1000km/s}
\end{center}
\end{figure}

\begin{figure}
\begin{center}
\includegraphics[width=\columnwidth]{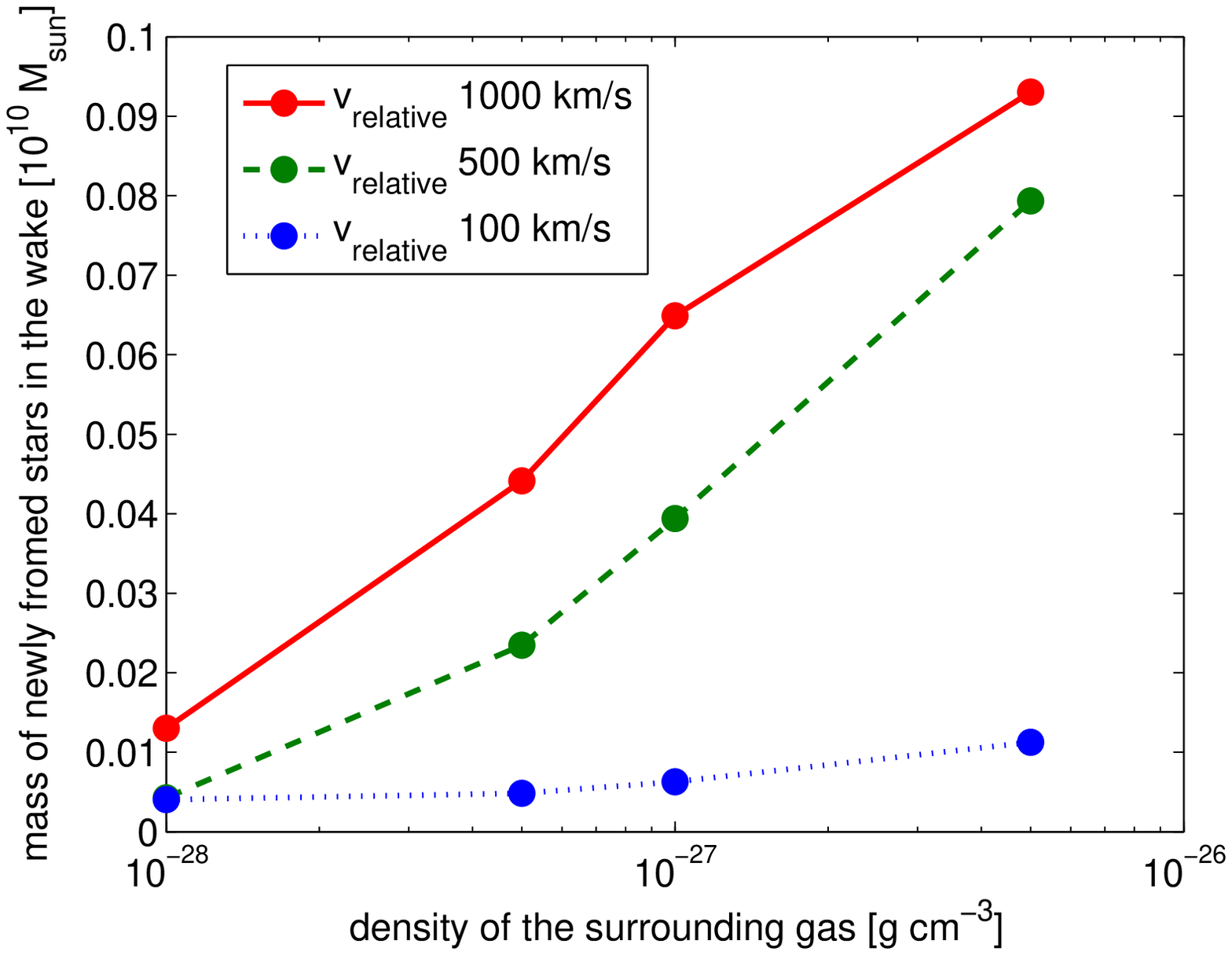}
\caption{Mass of newly formed stars in the wake (distances larger
than 10 kpc above the disc) after 500 Myr of evolution.}
\label{mass_newly_formed_stars}
\end{center}
\end{figure}

\begin{figure*}
\begin{center}
\includegraphics[width=\textwidth]{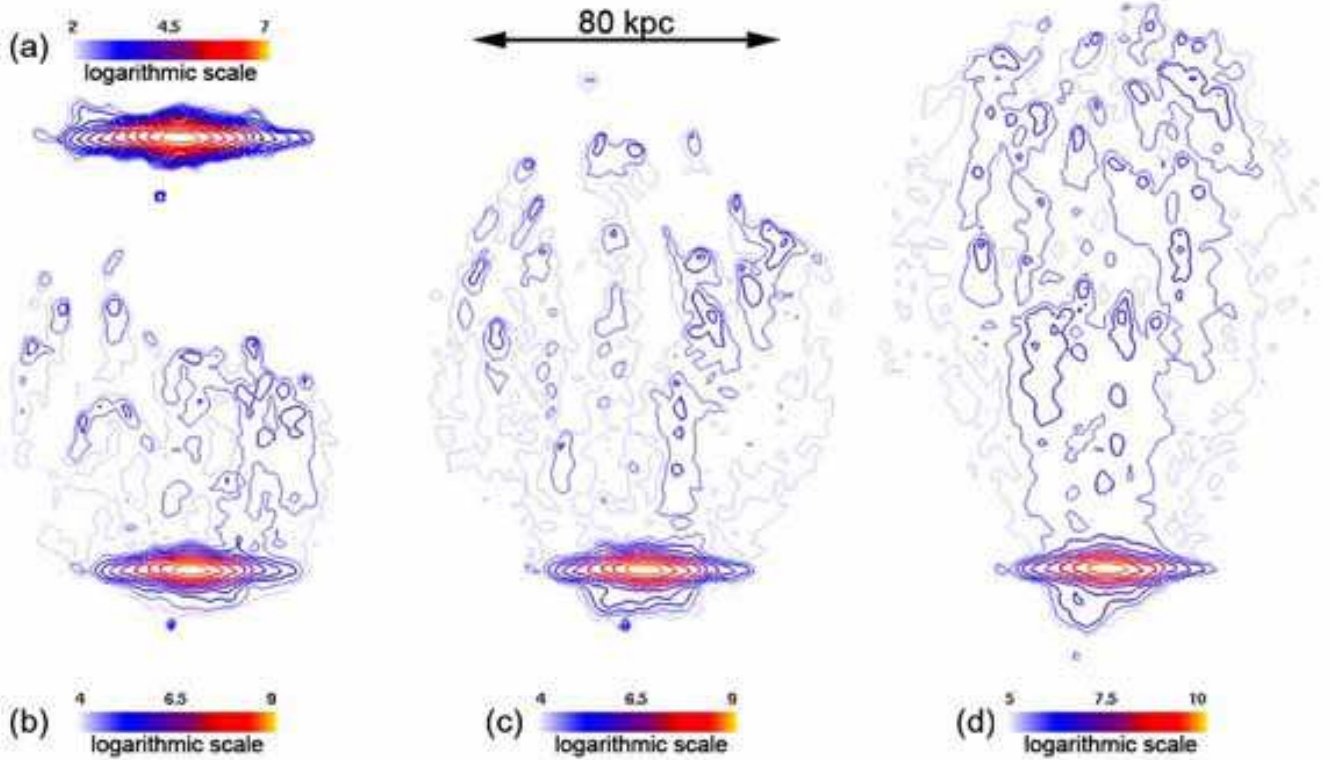}
\caption{The surface density of the total stellar component (in
relative units and a logarithmic scale) after 500 Myrs of ram
pressure acting on the galaxy. The relative velocity is 500 km/s for
all shown cases, whereas the surrounding density is varied((a)
$1\times10^{-28}$ g/cm$^{3}$, (b) $5\times10^{-28}$ g/cm$^{3}$, (c)
$1\times10^{-27}$ g/cm$^{3}$, (d) $5\times10^{-27}$ g/cm$^{3}$).
Note that the panels have different scales (see colour bars).}
\label{2d_stellar_comp_500_kms}
\end{center}
\end{figure*}

\begin{figure*}
\begin{center}
\includegraphics[width=\textwidth]{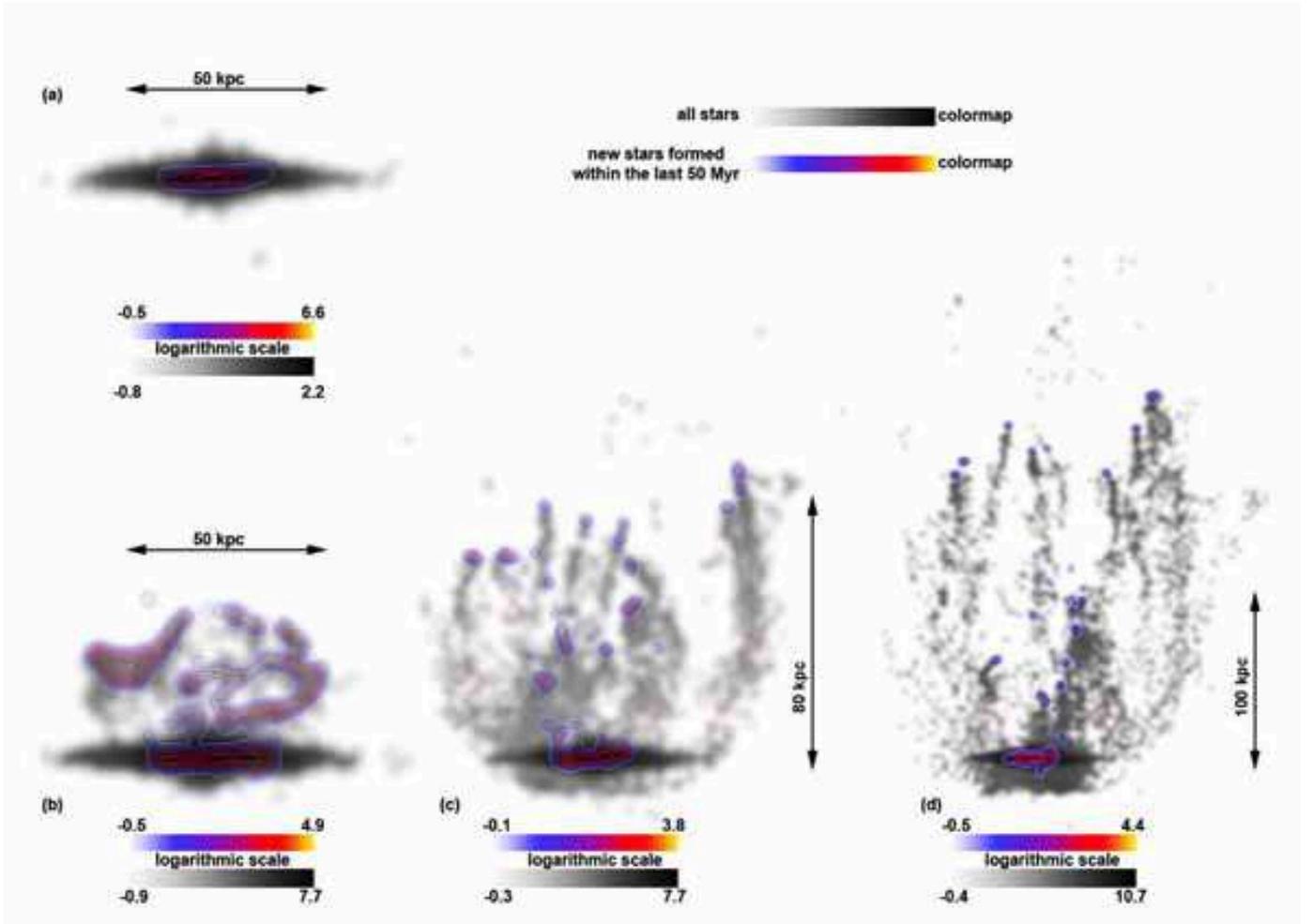}
\caption{The surface density of the total stellar component (in grey
scale) and newly formed stars within the last 50 Myr (isolines) for
different timesteps ((a) 50 Myr, (b) 250 Myr, (c) 500 Myr and (c)
750) of run 1. The relative velocity is 1000 km/s and the
surrounding gas density is $1\times10^{-28}$ g/cm$^{3}$. Note that
the panels have different scales in the density and length scales.}
\label{all_stars_sim_1}
\end{center}
\end{figure*}

\begin{figure*}
\begin{center}
\includegraphics[width=\textwidth]{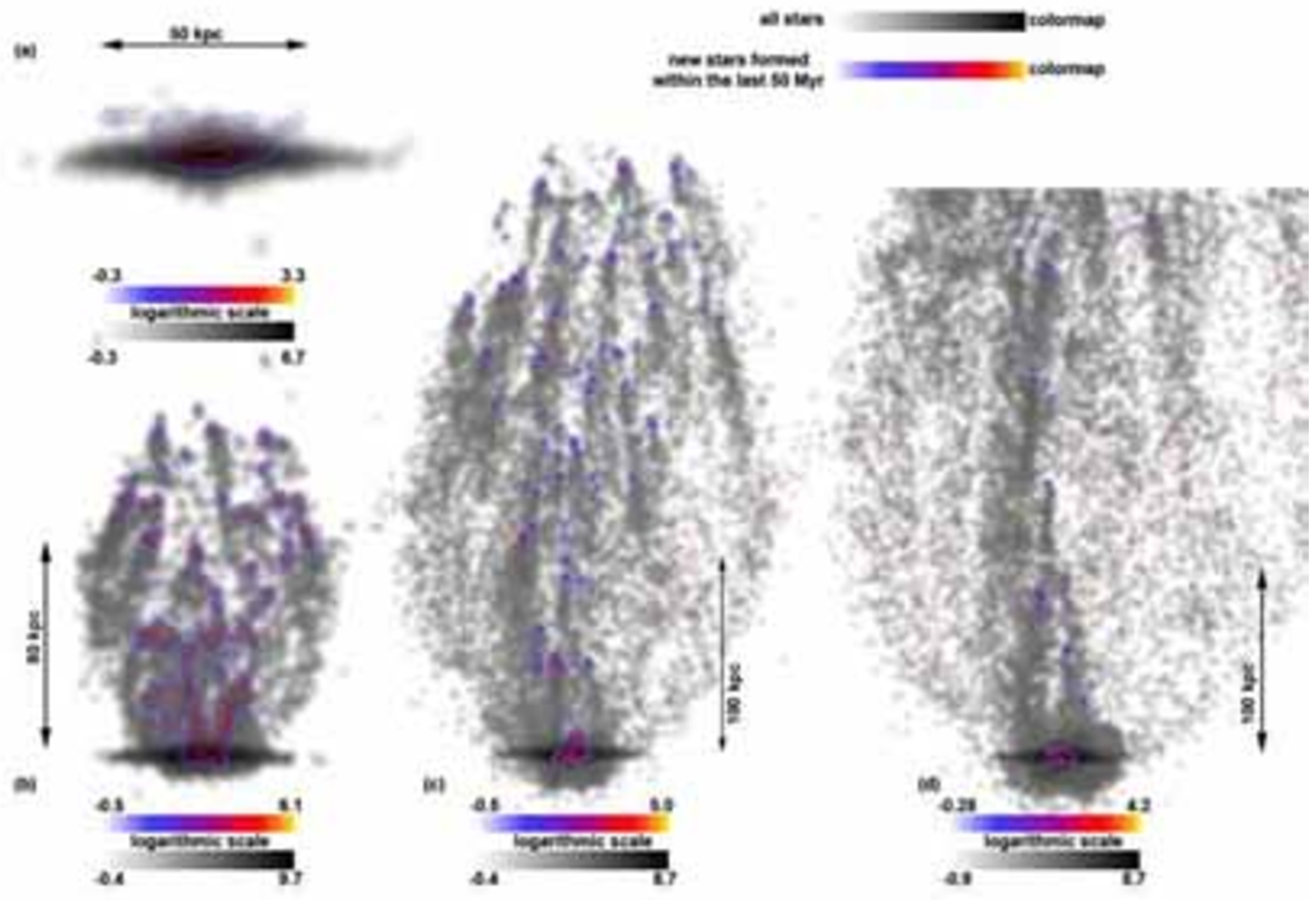}
\caption{The surface density of the total stellar component (in grey
scale) and newly formed stars within the last 50 Myr (isolines) for
different timesteps ((a) 50 Myr, (b) 250 Myr, (c) 500 Myr and (c)
750) in simulation 2. The relative velocity is 1000 km/s and the
surrounding density is $5\times10^{-28}$ g/cm$^{3}$. Note that the
panels have different scales in the density and length scales.}
\label{all_stars_sim_2}
\end{center}
\end{figure*}

\begin{figure*}
\begin{center}
\includegraphics[width=\textwidth]{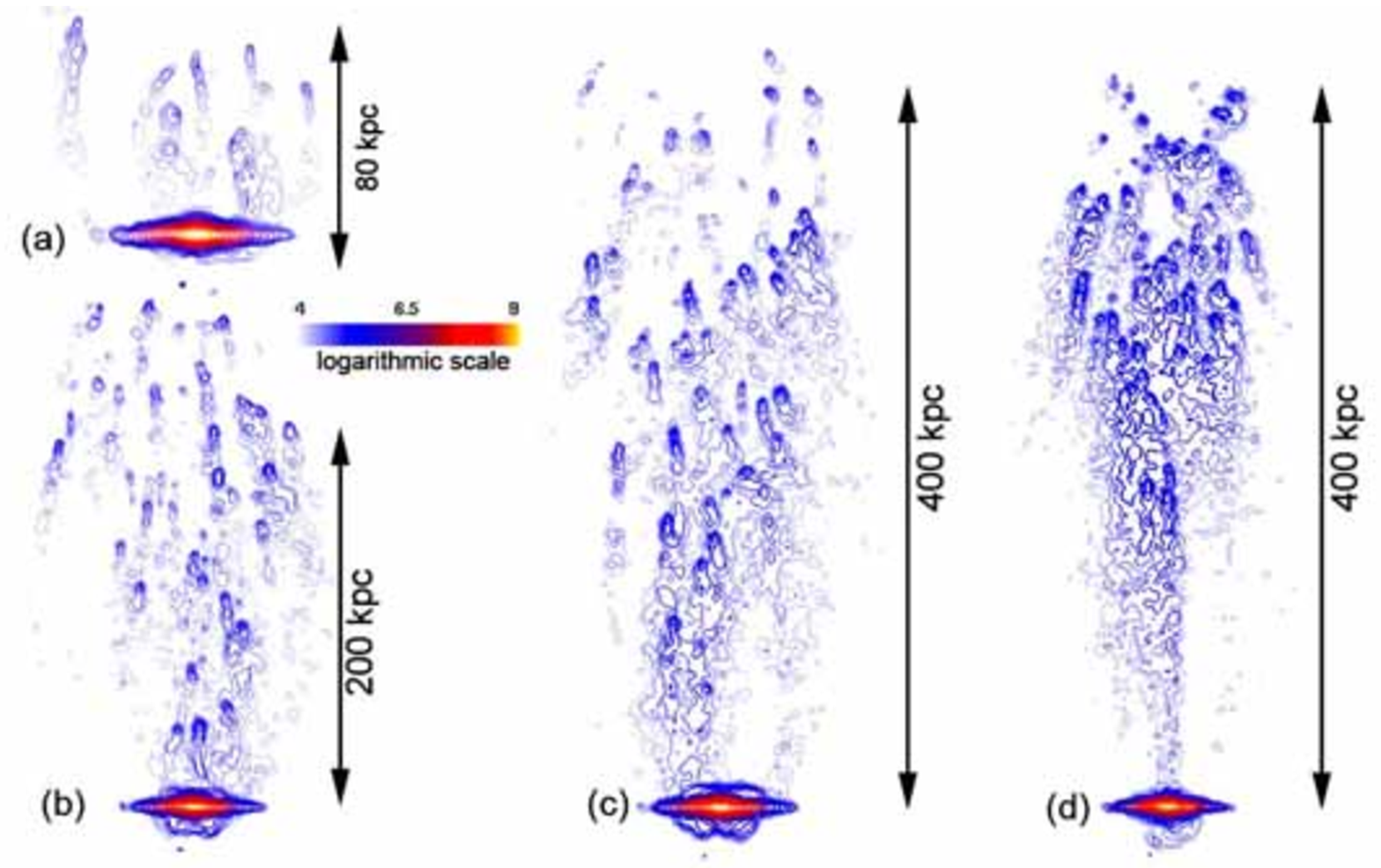}
\caption{The surface density of the total stellar component (in
relative units and a logarithmic scale) after 500 Myrs of ram
pressure acting face on. The relative velocity is for all shown
cases 1000 km/s, whereas the surrounding density is changed ((a)
$1\times10^{-28}$ g/cm$^{3}$, (b) $5\times10^{-28}$ g/cm$^{3}$, (c)
$1\times10^{-27}$ g/cm$^{3}$, (d) $5\times10^{-27}$ g/cm$^{3}$)}
\label{2d_stellar_comp_1000_kms}
\end{center}
\end{figure*}

Recent numerical studies (Kronberger et al. 2008) showed that ram
pressure does enhance the star formation in a galaxy. In order to
study a relation between the star formation enhancement and the
external ram pressure, we calculated the ratio between the temporal
integrated-star formation rate of each simulation including an
external ram pressure to the case of the isolated galaxy (i.e. no
external ram-pressure).\\
We vary the relative velocities and surrounding gas densities, as
summarised in Table \ref{ram_pressure_table}. The higher the
external density, the stronger the star formation enhancement (see
Fig. \ref{ratio_star_formation_vacuum}). The effect of the relative
velocity between the galaxy and the surrounding medium is relatively
small. The rate of star-formation enhancement changes depending on
the ambient pressure, which depends on the gas density. The rate
does nearly not change in the simulations with v$_{\rm{rel}}$=100
km/s. The reason is the depletion of the gas disc, which can be
stripped completely in the case of simulation 4, and the resulting
suppression of star formation in the disc. The faster the relative
velocity, the larger is the amount of stripped ISM, which will be
kinematically stronger disturbed, i.e. kinetic energy will be partly
converted in internal - thermal - energy, as in the case with lower
relative velocity. This leads to
fewer star forming regions than in cases with lower relative velocities.\\
From observations (e.g. Mihos et al. 2005) it is evident that an
intra-cluster stellar population exists. Some observations claim
that 10-40\% of all stars in galaxy clusters are in the space
between the galaxies (e.g. Feldmeier et al. 1998, Theuns \& Warren
1997, Arnaboldi et al. 2003). From recent numerical studies
(Kapferer et al. 2008) we know that ram pressure can cause star
formation in the stripped wake of galaxies Therefore we investigated
the amount and location of newly formed stars for the different
ram-pressure scenarios.\\
We found a relation between the relative velocity and the amount of
star forming regions in the wake. The higher the relative velocity
between the galaxy and the surrounding medium, the more star forming
region are developing in the wake. The explanation is the fact, that
stronger external ram pressure strips off more gas from the disc,
which leads to a depletion of gas in the galaxy. The highest ratio,
i.e. nearly all new stars being formed in the wake, can be found in
the simulation with the highest surrounding gas density and the
highest relative
velocity.\\
In the case of the lowest relative velocity (i.e. 100 km/s) almost
no new stars are formed in the wake, because nearly no gas is
stripped in these scenarios. Note that the wake is defined as region
above 10 kpc the stellar disc (i.e. the same definition as in
Kronberger et al. 2008).\\
In Fig. \ref{newly_formed_stars_in_wake_to_total_500km/s} and Fig.
\ref{newly_formed_stars_in_wake_to_total_1000km/s} the evolution of
the amount of newly formed stars in the wake relative to the total
amount of newly formed stars is shown. In the case of 1000 km/s
relative velocity and a surrounding gas density of $5\times10^{-27}$
g/cm$^{3}$ more than 95\% of all new stars are formed in the wake,
as the disc as stripped almost completely. In the scenarios with a
relative velocity of 500 km/s, the amount decreases to nearly 70\%
in the case with the highest surrounding gas density.
Note that the ratio of star formation in the disc to the wake
saturates at different timescales, depending on the strength of the
ram pressure. We find timescales in the range from several 50 Myrs
in the strongest ram pressure cases to nearly 300 Myrs in the
scenarios with the lowest surrounding gas densities. This behaviour
reflects the ram-pressure stripping time scales, i.e. the time until
the ram pressure has stripped the gaseous disc to a radius of
equilibrium between internal and external forces, i.e. the
stripping radius.\\
As a next step we investigated the amount of newly formed stars in
the stripped wake as a function of the ram pressure. In Fig.
\ref{mass_newly_formed_stars} the total mass of newly formed stars
as a function of the surrounding gas density is shown. It is
remarkable that nearly $1\times10^{9}$ M$_{\sun}$ of stellar mass is
present in the wake after 500 Myrs of evolution in the strongest ram
pressure scenario. Even in the case of a very low relative velocity
of 100 km/s $1\times10^{8}$ M$_{\sun}$ of stellar matter can be
found at distances larger than 10 kpc to the disc.\\
How the morphology of the stellar distribution is affected by
different strengths of ram pressure can be seen in Fig.
\ref{2d_stellar_comp_500_kms} and Fig.
\ref{2d_stellar_comp_1000_kms}, respectively. The two figures show
the stellar distribution after 500 Myrs of ram pressure acting on
the galaxy. The isocontours show the projected stellar density for
the different ram-pressure scenarios. The surface density varies by
five orders of magnitudes, with the maximum at the very centre of
the disc. The stronger the ram pressure the more gas is stripped
from the galaxies (see next section for details) which acts as gas
reservoir for new stars. Assuming a constant mass to light ratio
these two figures can be directly interpreted as optical
observations. In the cases of v$_{\rm{rel}}$=500 km/s three times
less luminous stellar regions than in the central parts are seen up
to 120 kpc behind the disc (see Fig. \ref{2d_stellar_comp_500_kms}).
In the cases with v$_{\rm{rel}}$=1000 km/s the distances can reach
up to 400 kpc. In the case highlighted as (d) in Fig.
\ref{2d_stellar_comp_1000_kms} a stellar 'highway' with luminosities
four to five orders of magnitudes lower than in the central disc are
present.\\
The distribution of newly formed stars within the last 50 Myr can be
seen in Fig. \ref{all_stars_sim_1} and Fig. \ref{all_stars_sim_2},
respectively. The two simulations, i.e. run 1 and 2, are shown here
in an exemplary way. The chosen time interval of 50 Myr is the
approximate life time of less massive B stars in OB-associations.
These associations would be observable in H$\alpha$. In Fig.
\ref{all_stars_sim_1} the distribution of all stars (grey scale) and
the young stars (formed within the last 50 Myr; isolines) are shown
in a logarithmic scale. The timesteps (a), (b), (c) and (d)
correspond to 50 Myr, 250 Myr, 500 Myr and 750 Myr after the ram
pressure started to act on the galaxy face on. These young stellar
components can be seen throughout the simulation in both components,
in the remaining gaseous disc and in the stripped wake.
In simulation 2 (Fig. \ref{all_stars_sim_2}) the same trend can be
seen. The stronger the ram pressure, the smaller the star forming
regions in the remaining disc. In this case more OB-association like
regions can be found in the wake. One important conclusion is, that
an overall high star formation rate does not imply high H$\alpha$
fluxes originating from the disc, but more H$\alpha$ flux observable
in the wake. But as the surface brightness drops one would observe
less star forming systems due to observational limitations, although
more stars are formed compared to the isolated case.
Simulations 1 and 2 can be related as low ram-pressure scenarios in
comparison to the strength of ram pressure expected in massive
galaxy clusters. Nevertheless the relatively low ram pressure
changes the morphology of the galaxy on very short timescales (less
than 250 Myr). From an observational point of view the system would
change from a star-forming, blue system, to a red, passively
evolving system, although stronger star formation is present as in
the isolated scenario. Therefore we can conclude here, that ram
pressure enhances the overall star formation of disc galaxies and
alters their star formation morphology locally, from star
forming discs to star forming wakes.\\
In addition to the stellar components at large distances from the
disc in the stripped wake an additional stellar component is present
in front of the disc, opposite to the stripped wake. The stars
originate from the gaseous wake in which they were originally
formed. As the stars do not feel the ram pressure of the surrounding
gas after they have formed, gravity is the the only force they are
exposed to. Therefore they will free-fall to the gravitational
centre, which is the disc embedded in the dark-matter halo. As the
interaction is collisionless, stars will pass through the centre and
appear on the other side and oscillate around the centre of the disc
leading to an observable component in the opposite direction of the
ram pressure. The same behaviour can be observed in the stripped
gas, whenever gas is moving into the slipstream of the remaining gas
in the disc, it partly falls back onto the disc. More details on the
influence of ram pressure on the ISM is given in the next sections.

\subsection{The influence of ram pressure on the gaseous disc}
\begin{figure}
\begin{center}
\includegraphics[width=\columnwidth]{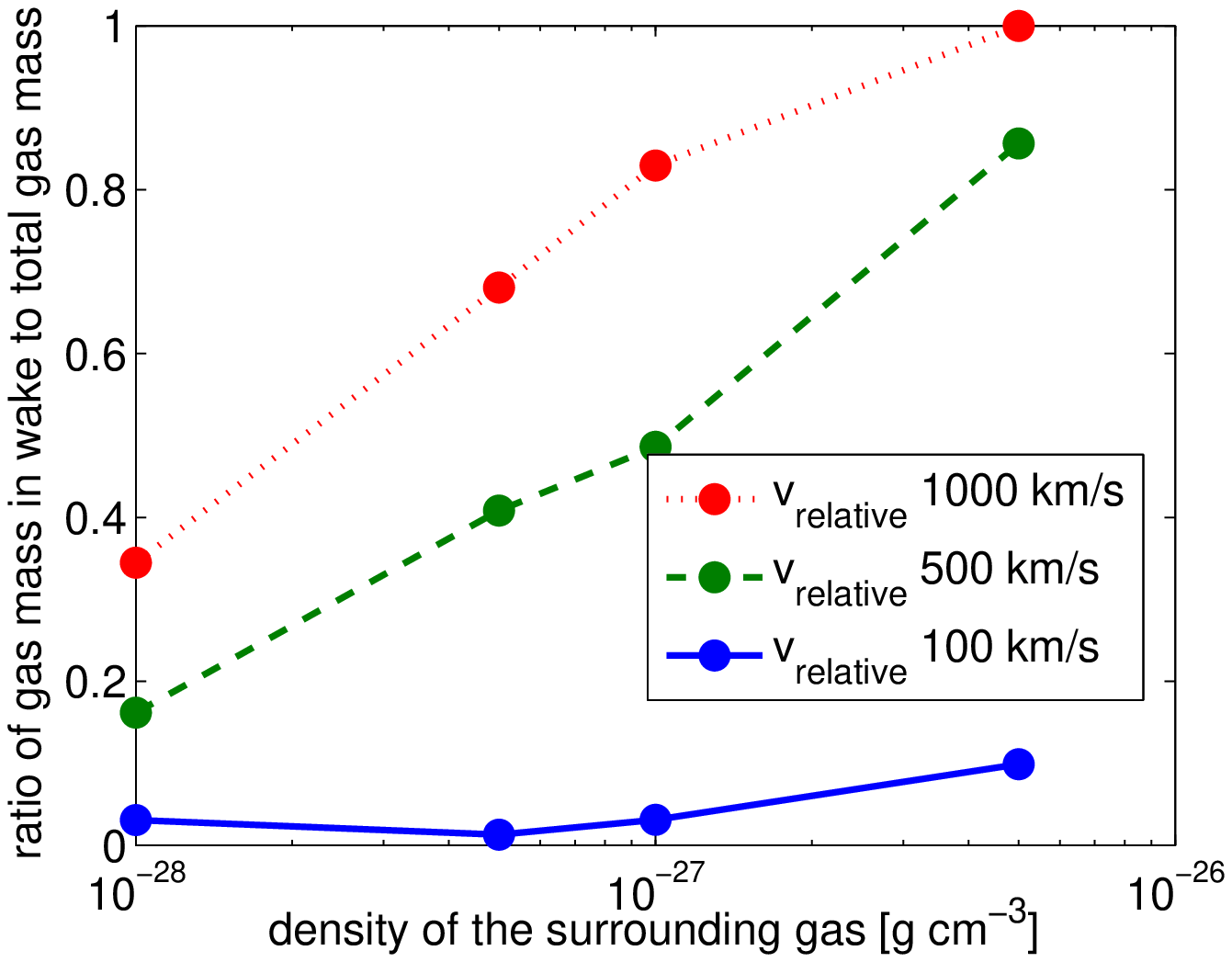}
\caption{Ratio of gas mass in the wake to the total gas mass of the
system (distances larger than 10 kpc behind the disc) after 500 Myr
of evolution for simulations with different densities and
v$_{\rm{rel}}$.} \label{ratio_gas_mass}
\end{center}
\end{figure}

\begin{figure}
\begin{center}
\includegraphics[width=\columnwidth]{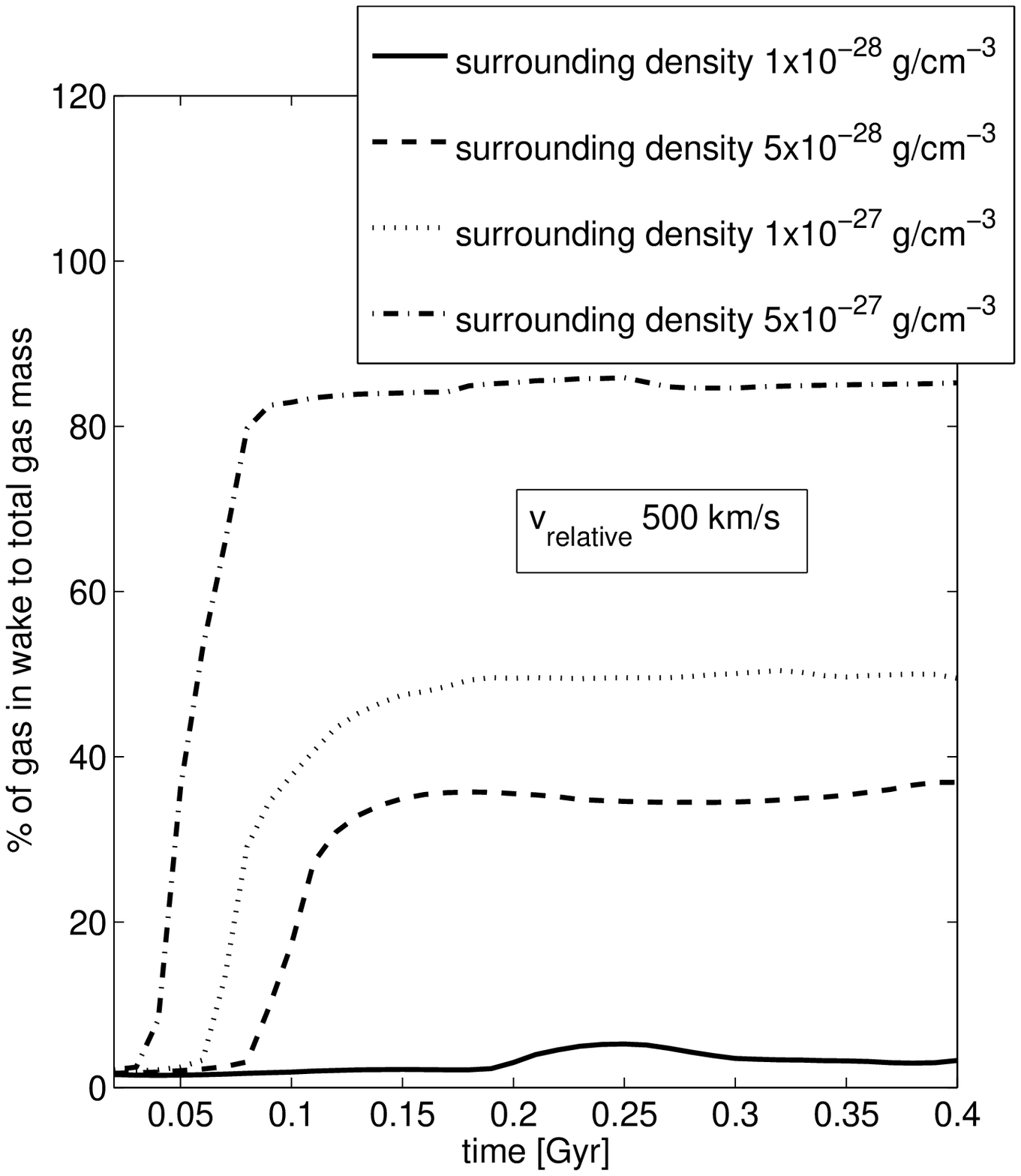}
\caption{Evolution of the gas mass in the wake to the total gas mass
(distances larger than 10 kpc above the disc).}
\label{gas_mass_wkae_evo_500}
\end{center}
\end{figure}

\begin{figure}
\begin{center}
\includegraphics[width=\columnwidth]{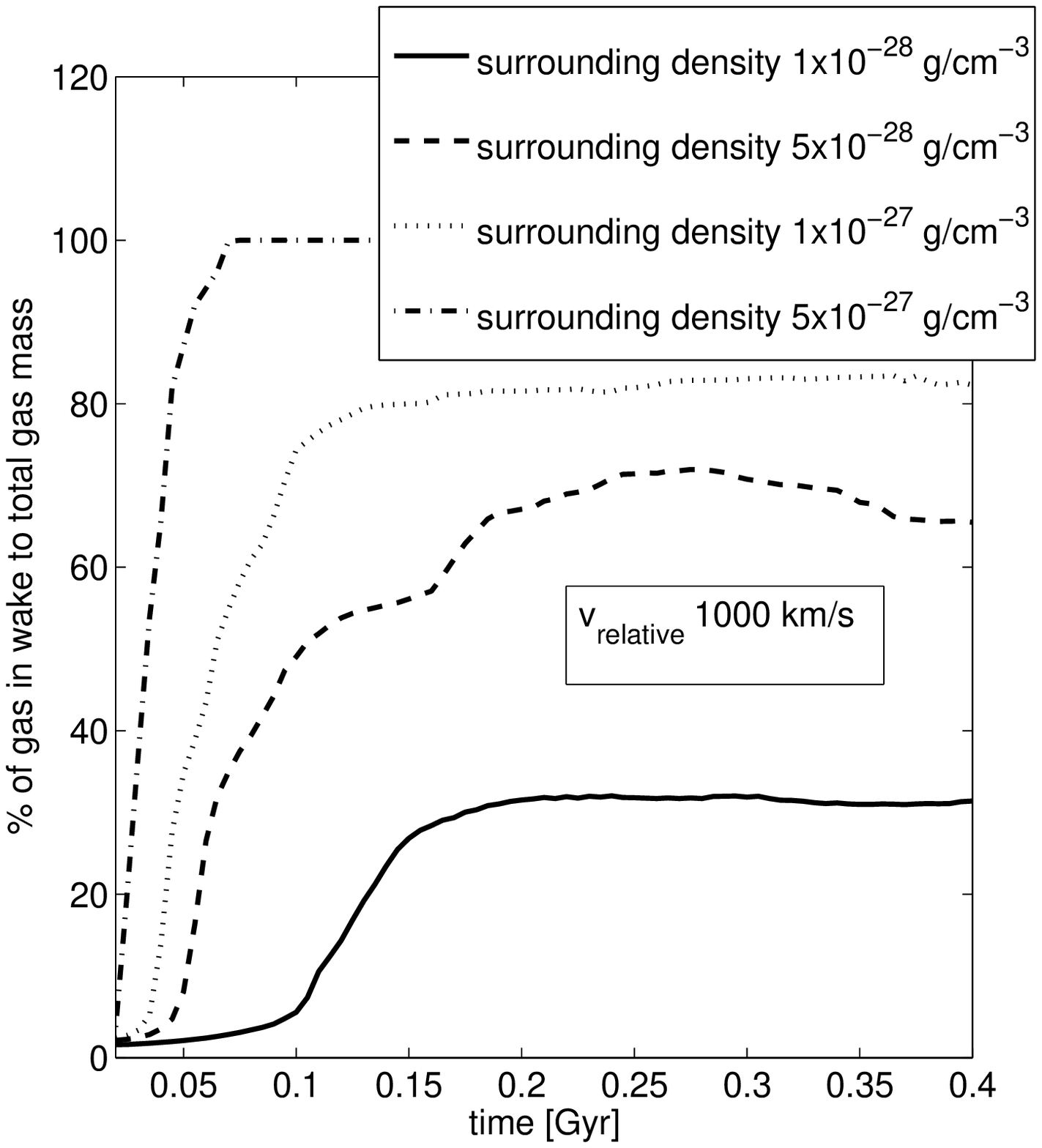}
\caption{Evolution of the gas mass in the wake to the total gas mass
(distances larger than 10 kpc above the disc).}
\label{gas_mass_wkae_evo_1000}
\end{center}
\end{figure}

How different strengths of ram pressures will affect the gaseous
disc of the galaxy will be investigated in this section. In Fig.
\ref{ratio_gas_mass} the ratio of the gas mass in the wake to the
total gas mass as a function of the surrounding gas density for the
three different relative velocities after 500 Myr of evolution is
shown. As the ram pressure depends on the product of the square of
the relative velocity and the surrounding gas density the increase
for the scenarios with v$_{\rm{rel}}$=1000 km/s and
v$_{\rm{rel}}$=500 km/s is stronger than the increase for
v$_{\rm{rel}}$=100 km/s.\\
Even low gas densities can deplete the gas mass in the disc. Up to
40\% of the gas mass in the disc is reduced in the case of the
lowest density of the surrounding gas and a relative velocity of
1000 km/s. If the gas density is larger than $5\times10^{-27}$
g/cm$^{3}$ the gas reduction in the disc is strong, the gaseous disc
vanishes in the scenario with 1000 km/s and only 20\% of the gaseous
disc survives in the scenario with 500 km/s. Note that even a very
low relative velocity of 100 km/s does reduce the gas disc by 10\%
in the case with $5\times10^{-27}$
g/cm$^{3}$ surrounding gas density.\\
The timescales of gas depletion in the disc are shown in Fig.
\ref{gas_mass_wkae_evo_500} and Fig. \ref{gas_mass_wkae_evo_1000},
respectively. The higher the ram pressure the faster an equilibrium
state of ram pressure and restoring, gravitational, forces is
established. The fastest gas depletion in the maximum ram pressure
scenario (v$_{\rm{rel}}$=1000 km/s and surrounding gas density of
$5\times10^{-27}$ g/cm$^{3}$) is reached after 100 Myr. It is
interesting that the equilibrium state is always reached in
timescales below 150 - 200 Myr. Only in the scenario with 500 km/s
relative velocity and the lowest surrounding gas density
$1\times10^{-28}$ g/cm$^{3}$ does it take 250 Myr to reach the
maximum, but as a large amount of gas falls back to the disc the
amount of stripped gas is less. Again the effect of a slipstream
caused by the the remaining gas disc can be observed.
The same effect can be seen in the simulation with 100 km/s relative
velocity and a surrounding gas density of $5\times10^{-28}$
g/cm$^{3}$.

\subsection{The distribution of the different components}
\begin{figure*}
\begin{center}
\includegraphics[width=\textwidth]{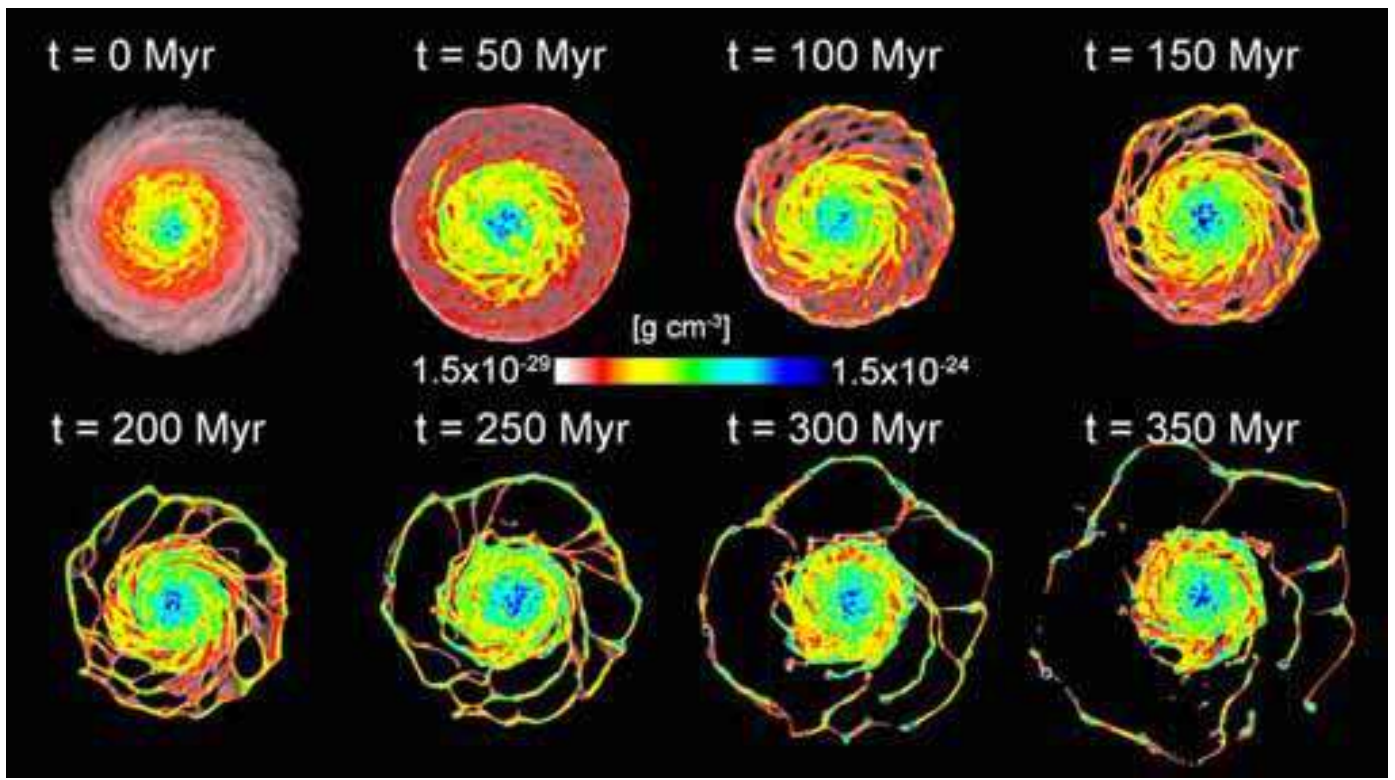}
\caption{Gas density of the total gas in the disc for different
timesteps for run 1. The external ram pressure is $1\times10^{-12}$
dyn/cm$^2$.} \label{spiral_pattern_dmw2}
\end{center}
\end{figure*}

\begin{figure}
\begin{center}
\includegraphics[width=\columnwidth]{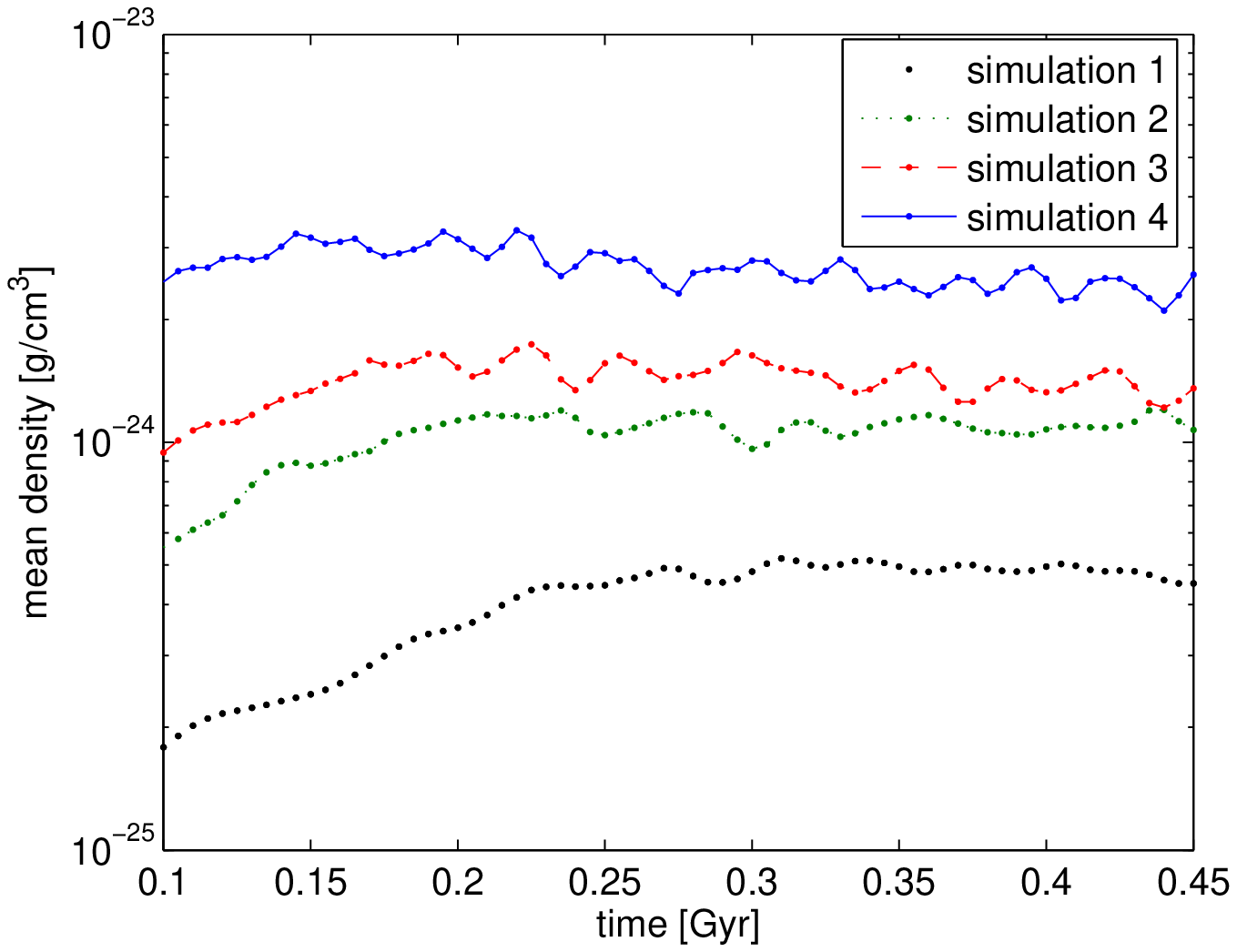}
\caption{Evolution of the mean local density of the cold
(T$<10^{6}$K) gas in the wake for simulations 1 to 4 (relative
velocity 1000 km/s and surrounding gas density in the range from
$1\times10^{-28}$ to $5\times10^{-27}$ g/cm$^{-3}$.}
\label{mean_density_evo}
\end{center}
\end{figure}

\begin{figure}
\begin{center}
\includegraphics[width=\columnwidth]{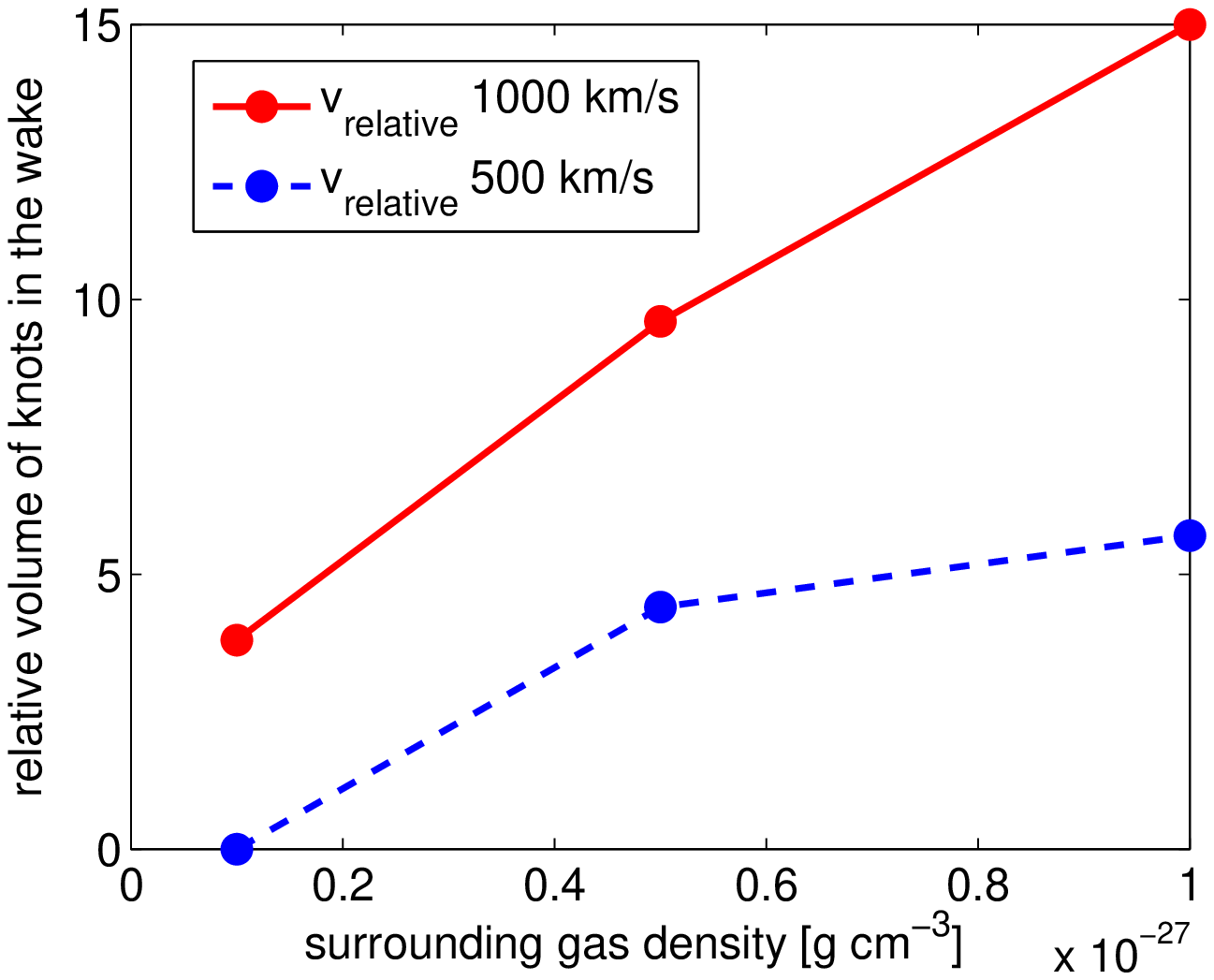}
\caption{Integrated volume of all isosurfaces with a fixed threshold
of $1\times10^{5}$ M$_{\sun}$/kpc$^{3}$ calculated from the three
dimensional gas density distribution in the wake as a function of
surrounding gas density for the scenarios with v$_{\rm{rel}}$=500
kms/s and v$_{\rm{rel}}$=1000 km/s.} \label{volume}
\end{center}
\end{figure}

\begin{figure*}
\begin{center}
\includegraphics[width=\textwidth]{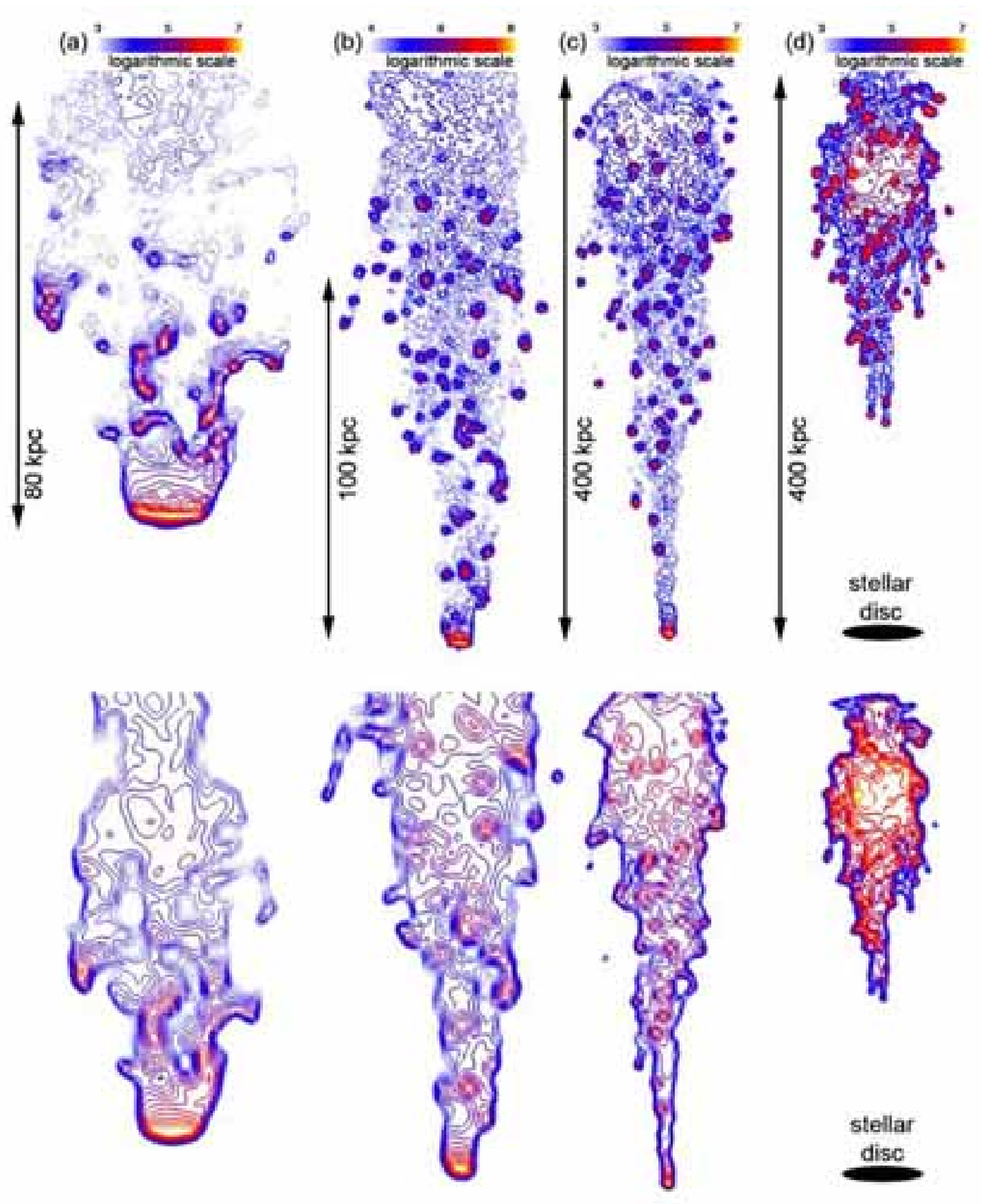}
\caption{Surface density isocontours of the total galaxies gas for
different surrounding gas densities ((a) $1\times10^{-28}$, (b)
$5\times10^{-28}$, (c) $1\times10^{-27}$ and (d) $5\times10^{-27}$ g
c$^{-3}$) and a relative velocity of 1000 km/s after 500 Myr of ram
pressure. The upper panels show a resolution of 3 kpc on a side for
a cell, whereas the lower panel gives the same quantity smoothed
with a Gaussian with $\sigma=4$ kpc to mimic a more realistic radio
telescope beam size. In the case of the maximum ram pressure - (d) -
the gaseous disc is completely stripped and therefore the location
of the old stellar disc is highlighted.} \label{gas_maps_1000_kms}
\end{center}
\end{figure*}

\begin{figure*}
\begin{center}
\includegraphics[width=\textwidth]{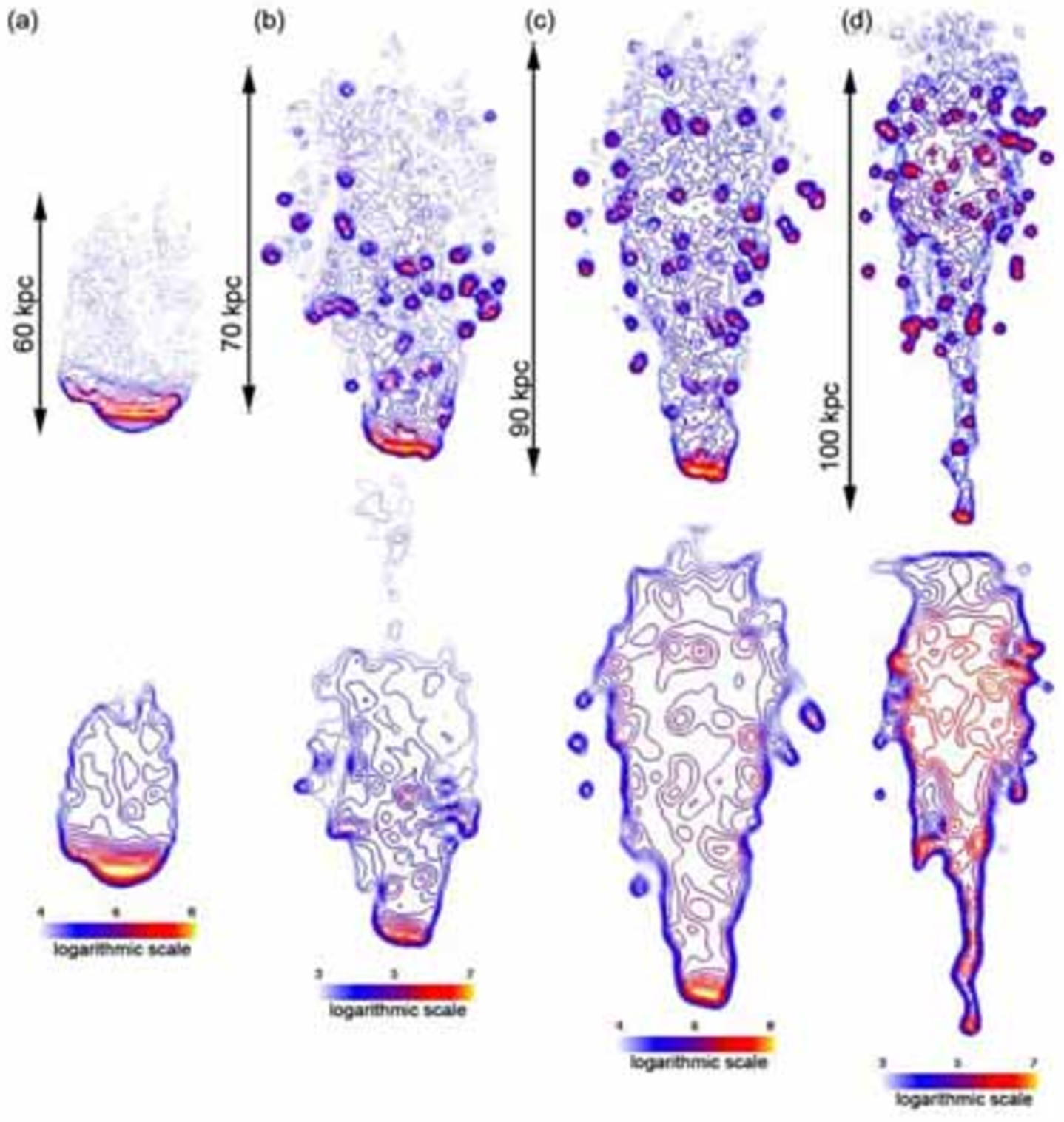}
\caption{Surface density isocontours of the total galaxies gas for
different surrounding gas densities ((a) $1\times10^{-28}$, (b)
$5\times10^{-28}$, (c) $1\times10^{-27}$ and (d) $5\times10^{-27}$ g
c$^{-3}$) and a relative velocity of 500 km/s after 500 Myr of ram
pressure. The upper panels show a resolution of 3 kpc on a side for
a cell, whereas the lower panel gives the same quantity smoothed
with a Gaussian with $\sigma=4$kpc to mimic a more realistic radio
telescope beam size.} \label{gas_maps_500_kms}
\end{center}
\end{figure*}

\begin{figure*}
\begin{center}
\includegraphics[width=\textwidth]{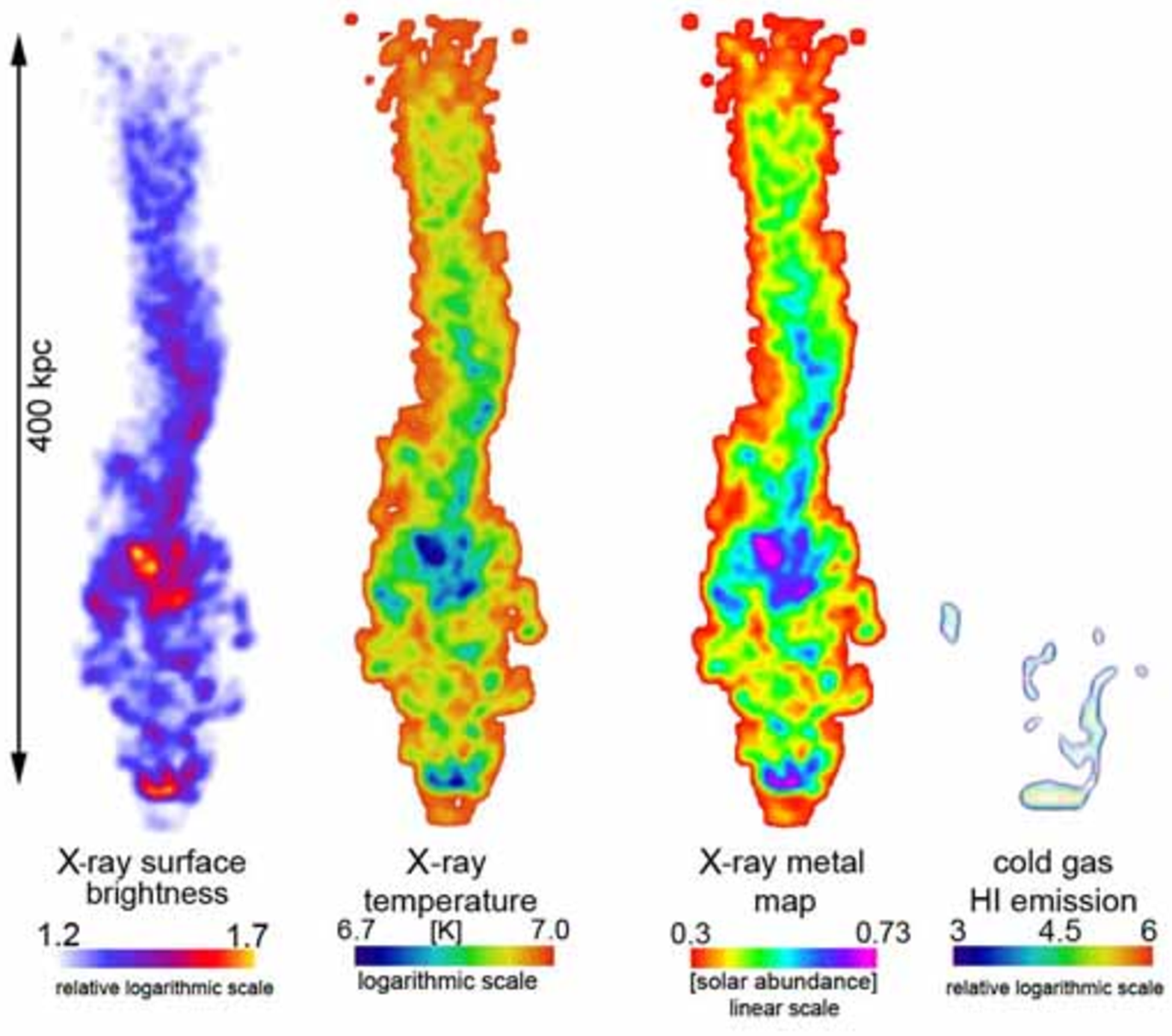}
\caption{Mock X-ray and HI observations for simulation 1, i.e. 1000
km/s relative velocity and a surrounding gas density of
$1\times10^{-28}$ g cm$^{-3}$. The ram pressure has acted for 500
Myr on the model galaxy.} \label{all_dmw_2_new}
\end{center}
\end{figure*}

\begin{figure*}
\begin{center}
\includegraphics[width=\textwidth]{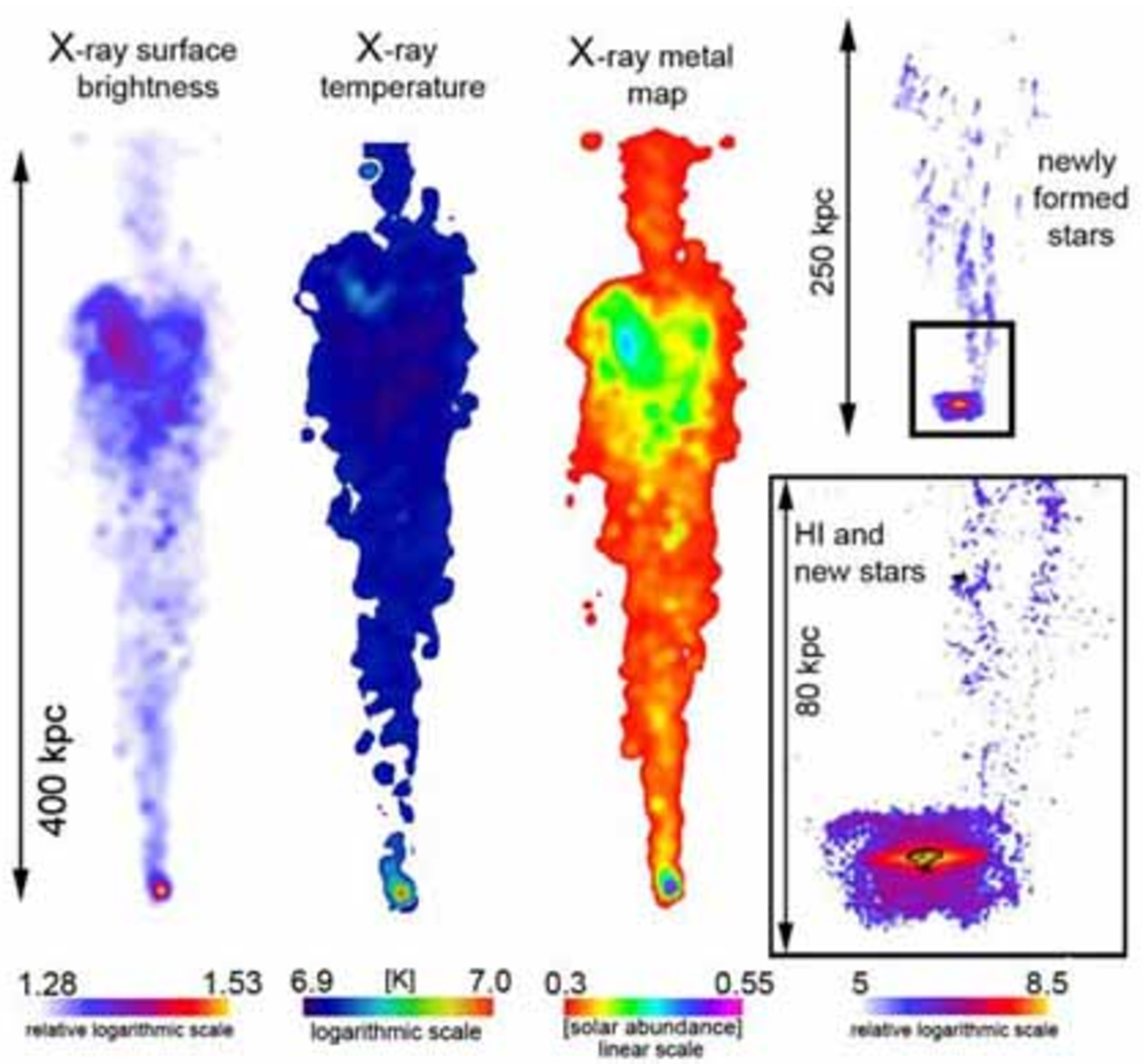}
\caption{Mock X-ray and HI observations for simulation 3, i.e. 1000
km/s relative velocity and a surrounding gas density of
$1\times10^{-27}$ g cm$^{-3}$. The ram pressure has acted 500 Myr on
the model galaxy.} \label{all_dmw_3_new}
\end{center}
\end{figure*}

\begin{figure*}
\begin{center}
\includegraphics[width=\textwidth]{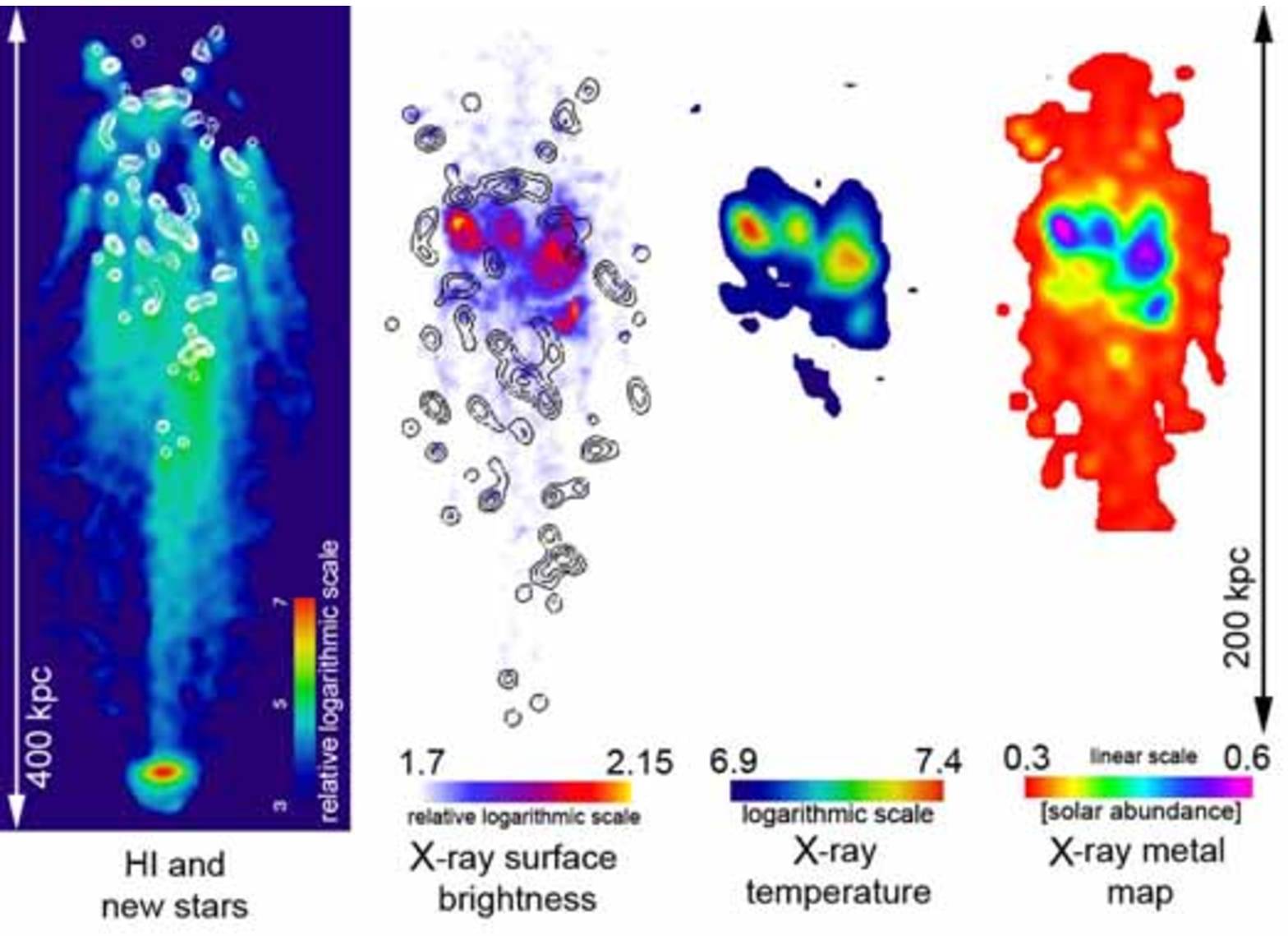}
\caption{Mock X-ray and HI observations for simulation 4, i.e. 1000
km/s relative velocity and a surrounding gas density of
$5\times10^{-27}$ g cm$^{-3}$. The ram pressure has acted 500 Myr on
the model galaxy, it is the simulation in which the gaseous disc as
completely stripped from the galaxy. X-ray temperature map in K.}
\label{all_dmw_5_new}
\end{center}
\end{figure*}

\begin{figure*}
\begin{center}
\includegraphics[width=\textwidth]{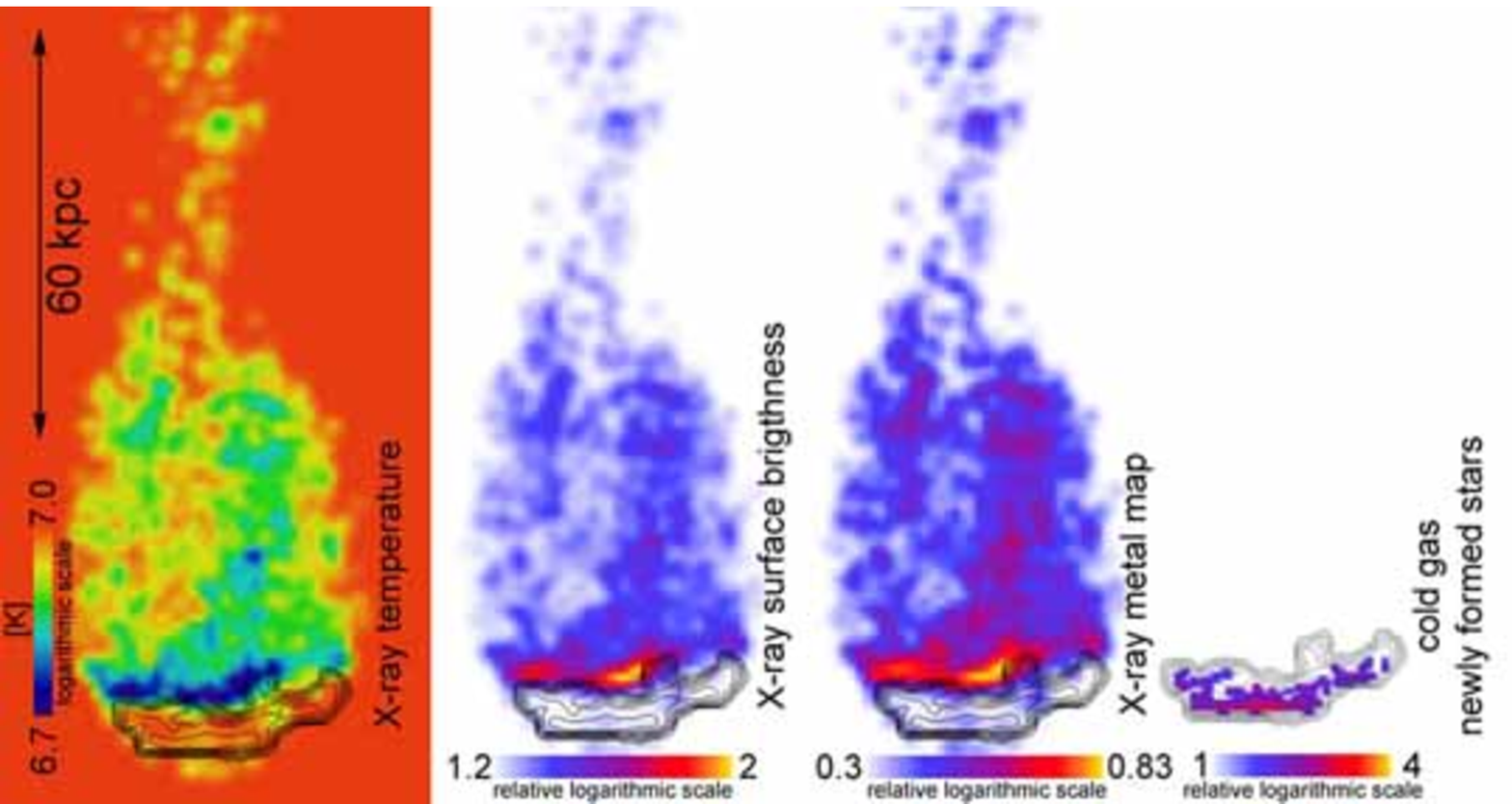}
\caption{Mock X-ray and HI observations and newly formed stars for
simulation 5, i.e. 500 km/s relative velocity and a surrounding gas
density of $1\times10^{-28}$ g cm$^{-3}$. The ram pressure has acted
for 500 Myr on the model galaxy, in this, the simulation with the
weakest stripping.} \label{all_dmw_6_new}
\end{center}
\end{figure*}

In this section it is investigated how ram pressure distributes the
gaseous component of the galaxy in the surrounding region. Therefore
we calculate for several simulations mock observations in the
different wavelengths,i.e. HI and X-ray observations. In Fig.
\ref{gas_maps_1000_kms} and Fig. \ref{gas_maps_500_kms} the
projected gas density of all the gas which belongs to the galaxy can
be seen. In the upper panel a resolution of 3 kpc is chosen. In the
lower panel the same map is smoothed with a Gaussian with $\sigma=4$
kpc to mimic a more realistic beam of a radio telescope for a galaxy
observed at a distance of the Virgo galaxy cluster. The cold ISM gas
(gas temperature T$<$20000 K), which would correspond to HI emission
is shown in the Figs. \ref{all_dmw_2_new}, \ref{all_dmw_3_new} and
\ref{all_dmw_5_new}. An interesting feature is the increase of gas
concentrations in the wake as a function of increasing ram pressure.
Note that the gas density projection is shown with isolines in a
logarithmic scale.\\
The gas knots in the wake are typically 2 to 3 orders of magnitude
less dense as the central parts of the gaseous disc. The origin of
these knots lies in the original gas distribution of the undisturbed
disc, i.e. the spiral arms.
\begin{figure*}
\begin{center}
\includegraphics[width=\textwidth]{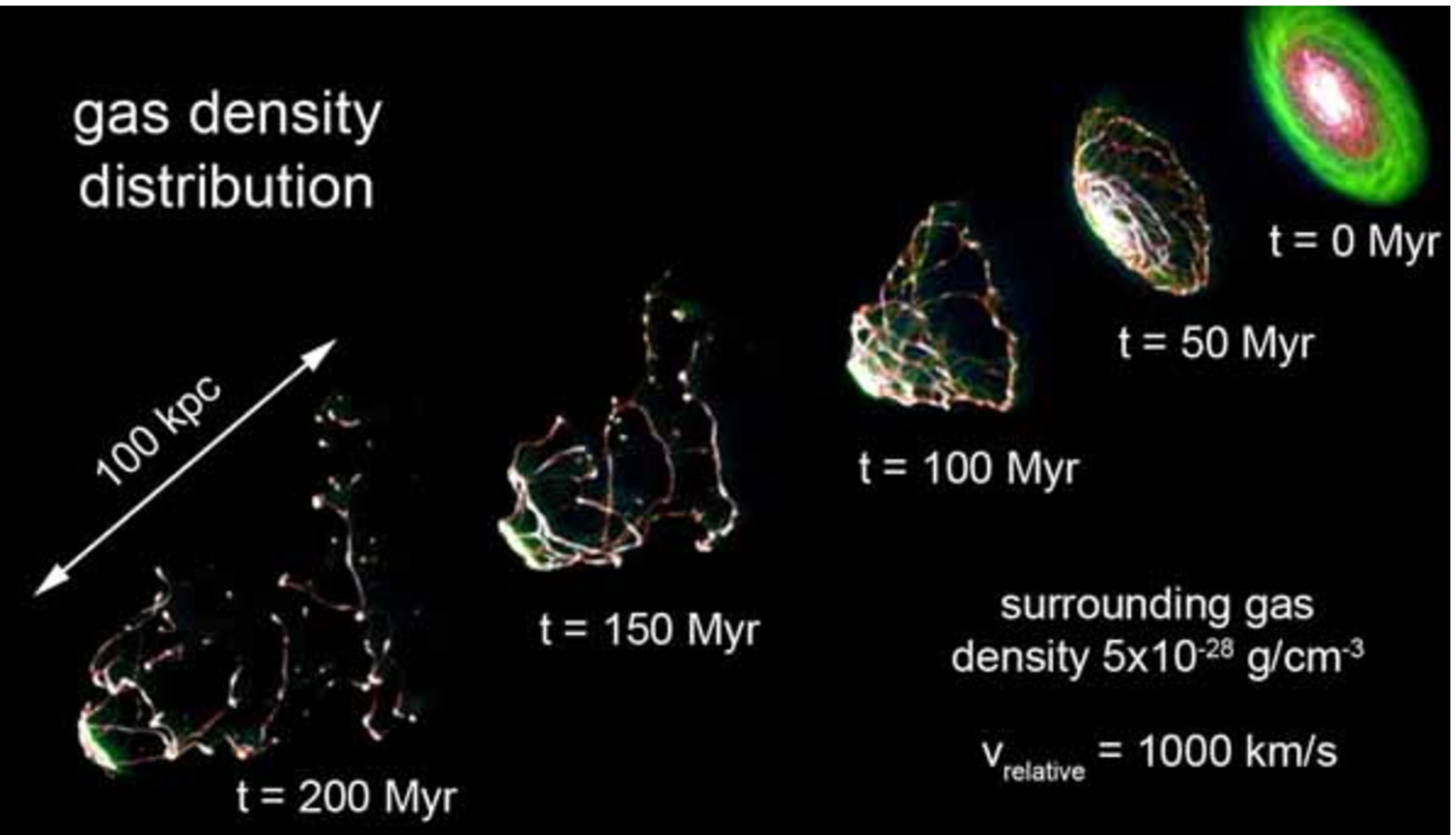}
\caption{The gas distribution for five different time steps, showing
the evolution of the gas component over 200 Myr. The ram pressure
and the relative velocity correspond to run 2. The colour map
indicates the density, i.e. white denser to green less dense
regions.} \label{gas_dens_dmw4_evo}
\end{center}
\end{figure*}
In Fig. \ref{gas_dens_dmw4_evo} the evolution of the gaseous spiral
arms for one simulation over 200 Myr of ram pressure is shown in
detail. As the disc rotates the ram pressure removes the complex gas
distribution in the disc. As the rotating spiral arms are removed
from the galaxy, selfgravitating structures of gas form from the
spiral arms, which evolve into the star-forming regions in the wake.
In Fig. \ref{spiral_pattern_dmw2} the gas density of the disc in run
1 (external ram pressure is $1\times10^{-12}$ dyn/cm$^2$) is shown
for 8 different timesteps. A very similar behaviour as reported by
Schultz and Struck (2001) can be seen, the compression of the
residual disc and the formation of flocculent spirals. The result is
an increase of the star formation in the residual gaseous disc.
If the ram pressure is increased the stripping radius moves inwards
the disc. This leads to the stripping of denser spiral arms, which
stay bound together longer and are less affected by the ram pressure
as the outer HI disc. In Fig. \ref{mean_density_evo} the evolution
of the mean local density of the gas at the position of the cold
(T$<10^{6}$K) gas, i.e. star forming regions, in the wake for
simulations 1 to 4 is shown. The rather constant mean local density
in the wake found in all simulations, shows that the stellar
feedback process does not destroy the gaseous knots in the wake.
They stay bound on very long timescales, more than several hundreds
of Myrs.
In Fig. \ref{volume} the integrated volume of all gas structures in
the wake with a isosurface density of $1\times10^{5}$
M$_{\sun}$/kpc$^{3}$ after 500 Myr of ram pressure is shown.
As the ram pressure increases the volume, i.e. the number of gas
complexes, increases. If the surrounding gas density is increased by
a factor of 2 the number of complexes is increased by a factor of
1.6 in the case of v$_{\rm{rel}}$=1000 km/s.
In scenarios with 1000 km/s the ram pressure shows strong influence
on the gas disc and the resulting gas wake includes many complex
features. In the case (d) in Fig. \ref{gas_maps_1000_kms} the ram
pressure has stripped off the whole gaseous disc. The stripped gas
lies more than 150 kpc behind the disc after 500 Myr of ram
pressure, leading to an elongated gas structure with several gaseous
clumps. In the scenarios with with v$_{\rm{rel}}$=500 km/s of
relative velocities the gaseous disc is not completely stripped.
Nevertheless the gas clumps themselves are affected by the ram
pressure, leading to structures with gaseous wakes, a kind of
self-similarity.\\
In order to calculate comparable maps with observations we extracted
mock X-ray and HI observations from our simulations. The X-ray
surface brightness, X-ray temperature and X-ray metal maps are
produced as described in detail in Kapferer et al. (2007). The
initial metallicities of the galaxy were chosen to be constant for
the gas in the disc and the surrounding gas, respectively. The
values are 1 in solar abundances for the ISM and 0.3 in solar
abundances for the hot, surrounding medium. Only gas with
temperatures above $5\times10^{6}$ K is taken into account for the
X-ray observations. For the HI mock observations we only take into
account the cold (T$<2\times10^5$ K) gas. In addition we smooth the
gas distribution with a Gaussian with $\sigma=4$ kpc to mimic a more
realistic radio telescope beam size for galaxies in the distance of
the Virgo cluster. The mock observations are shown in Fig.
\ref{all_dmw_2_new}, \ref{all_dmw_3_new} and \ref{all_dmw_5_new},
respectively.
In the case of simulation 1 (see Fig. \ref{all_dmw_2_new}) the
resulting mock maps show an extended X-ray emitting wake with
several hundred kpc extent. Roughly 20 kpc and 150 kpc behind the
stellar disc a peak in the X-ray maps is present. The latter is the
region with the highest density of hot gas originating from the
gaseous disc. In the X-ray temperature map a cool region of gas can
be found at the same position, the X-ray metal map shows there the
highest metallicities. The cold HI gas does not reach such large
distances from the stellar disc. Nevertheless cool gas can be found
in blobs at nearly 100 kpc behind the disc. The effect of ram
pressure can be seen in X-rays most dramatically, especially the
X-ray metal map reveals many
structures in the wake.\\
An increase of the ram pressure by one order of magnitude changes
the HI observation dramatically, see Fig. \ref{all_dmw_3_new}.
Nearly all cold gas is stripped, only a small fraction of cool gas
is left in the centre of the disc. In addition to the cool gas the
stellar surface brightness of newly formed stars is shown. The HI
mock observations show a feature in opposite direction to the wake.
The reason for the feature is cool gas, which was transported into
the slipstream of the gaseous disc and as the disc protects this
region from the external ram pressure the gas has fallen onto the
disc, leading to features opposite to the direction of the ram
pressure. These features are common throughout the simulations and
survive typically some tens of Myrs.\\
Similar to the case of simulation 1 an X-ray maximum in the surface
brightness, temperature and metal map is seen behind the disc in the
direction of the ram pressure. In addition a large region of X-ray
emitting gas is observable at more than 300 kpc behind the disc.
This region is very prominent in the X-ray surface brightness and
metal map. After 100 Myr of ram pressure the equilibrium state
between the ram pressure on the disc and the restoring gravitational
forces is nearly reached (see Fig. \ref{gas_mass_wkae_evo_1000}),
therefore most of the material can be found at larger distances
behind the disc, than in the ram pressure scenario of simulation 1.
As the main ram-pressure stripping event has happened more than 400
Myr before the mock observation, the gas has had already more time
to mix with the surrounding gas, leading to less pronounced X-ray
structures in the wake. In the right upper panel in Fig.
\ref{all_dmw_3_new} a stellar surface brightness up to a distance of
250 kpc is shown. The asymmetry of the newly formed stars directly
behind the disc originate from a recent stripping event of a gaseous
spiral. Note, only newly formed stars within the last 500 Myr are
shown. The stellar components reach over the whole wake and end at
the X-ray emitting region in a distance of 400 kpc.
The strongest ram pressure of all simulations changes the global
property of the X-ray emission to a great extent. The maximum at a
distance of 400 kpc marks the region at which the galaxy has passed
when the gaseous disc was completely stripped, it is enclosed by
cool gas. Many stars are formed at this location and as they fall
back to the disc, they assemble to 'highways' of stellar light.\\
The simulations with less strong ram-pressure show all a common
feature, a long X-ray tail. Although not much of the gas cold gas is
stripped off and the star formation takes place mainly in or very
near the disc. A several tens of kpc long X-ray emitting hot gas
tail develops. In Fig. \ref{all_dmw_6_new} mock X-ray and H$\alpha$
observations including isolines showing the distribution of the cold
gas phase are presented for simulation 5 (surrounding gas density of
$1\times10^{-28}$ g cm$^{-3}$ and a relative velocity of 500 km/s).
The newly formed stars are mainly concentrated in the compressed
residual disc. Like in the observations of Yoshida et al. (2002 and
2004) some H$\alpha$ emitting gas can be found several kpc behind
the ram pressure affected disc in the forming gaseous wake.

\subsection{Comparison to other simulations of ram-pressure stripping}

Many of the results presented here are in good agreement with
previously published results in other simulations of ram pressure
stripping. In the literature (e.g. Roediger \& Brueggen 2007,
Vollmer et al 2001) the size of the unstripped disc is often
compared to the analytic estimate of Gunn \& Gott (1972). This
provides a convenient means of comparing simulations with different
initial disc structures. We therefore checked the stripping radius
for the simulations with a residual gaseous disc and found that the
stripping radius lies within 10 \% of the predicted radius (Gunn \&
Gott 1972, Domainko et al. 2006). As found in Vollmer et al. (2001)
a considerable fraction of the stripped gas can fall back on to the
disc. In our simulations this is due to the slipstream of the
residual gaseous disc. As in the simulations of Schulz \& Struck
(2001) the simulation presented here includes radiative cooling. The
effect of the cooling is quite strong in the wake of the stripped
gas, leading to lower temperatures and more dense regions, which
will then form stars. The time scales for stripping are comparable
to those found in Roediger et al. (2006) in the range of 100 Myr or
below and the overall sizes of the gaseous wakes comparable to those
presented in Roediger et al. (2006). As the simulations presented in
this work include feedback od star formation on the surrounding gas,
the resulting properties of the gaseous wake are not as directly
comparable to other simulations. A common difference is visible in
the structure of the stripped wake, due to the high resolution in
the presented simulations and the presence of spiral arms, the
stripped gas has many small structures, which are stripped, evolve
and survive in the wake over many hundreds of Myrs. In J{\'a}chym et
al.(2007) for an example the stripped ISM does not show clumps of
cool gas. In the eulerian grid simulations of Roediger et al. (2006)
the wake does show cooler, dense regions, but the number is much
less as in the simulations presented here. A possible explanation is
the lack of radiative cooling in the simulations by Jachym et al.
(2007) and Reodiger et al (2006).

\section{Comparison with observations}
Apart from the evidence for the effect of ram pressure on the
gaseous component of galaxies seen in the radio wavelengths, recent
observations in other wavebands have revealed the influence of ram
pressure on the stellar component. Yoshida et al. (2008) report on
unusual complex narrow blue and H$\alpha$ emitting filaments and
clouds extending up to 80 kpc south from an E+A galaxy RB199 in the
Coma galaxy cluster. The galaxy seems to be a galaxy-galaxy merger.
They conclude that the most plausible formation mechanism for these
star forming regions is ram-pressure stripping.\\
Another example how ram pressure stripping influences the gaseous
disc and the stripped material is presented in Yoshida et al. (2004,
2002). They investigated NGC 4388, a Seyfert 2 Virgo cluster galaxy,
with deep optical spectroscopy. Many H$\alpha$ regions in several
tens of kpc in the stripped wake were found. One explanation for the
off-disc H$\alpha$ emission is the Seyfert nucleus, acting as a
ionization source. Another mechanism is shocks introduced by the ram
pressure, which ionizes the cloud complexes. Some of the gas clouds
found have masses in the range of $10^5$ to $10^6$ M$_{\sun}$ and
sizes up to several hundred pc. The excitation mechanism in the
outer region of the stripped cloud complexes (r$>$12 kpc) is yet
unclear. In the simulations presented here, several comparable star
forming knots were found in the stripped wake. As in these gas
complexes star formation is present, one explanation for the
observed H$\alpha$ emission in the gas complexes far from the
Seyfert nucleus in NGC 4388 is ongoing star formation, comparable to
the mock observations of simulation 5
presented in Fig.\ref{all_dmw_6_new}.\\
Another striking observation is presented by Sun et al. (2007). They
find a 40 kpc H$\alpha$ tail and at least 29 emission-line objects
downstream a star-forming galaxy in the the galaxy cluster A3627.
The galaxy has in addition to the H$\alpha$ tail an X-ray tail with
a length of 70 kpc, very similar to the stripping situation in shown
in Fig. \ref{all_dmw_3_new}. Both can be found in our simulations
as well.\\
The longest X-ray tail formed from stripped gas originating from a
galaxies observed so far is located in the Virgo galaxy cluster
(Randall et al. 2008), a Chandra observation of M86 in Virgo and the
surrounding field. The X-ray tail is 150 kpc in projection and a
simple estimate of the true trajectory yields an X-ray tail with a
length of 380 kpc. In addition they find an X-ray plume behind the
X-ray emission of M86, very similar to our finding that the first
X-ray maximum can be found behind the ram-pressure affected disc, as
a result of the first compressional heating of the
stripped gas.\\
Recently Kim et al. (2008) presented Chandra and XMM-Newton
observations of a prominent X-ray tail of the galaxy NGC 7619. They
found in the tail a significantly higher metallicity than in the
surrounding regions, and conclude that this indicates, that the gas
originates from the galaxy. In addition they find discontinuities in
the X-ray surface brightness map in the opposite direction, which
suggest a supersonic relative velocity of the galaxy with respect to
the surrounding gas.\\
The origin of intergalactic HI complexes is still under debate. A
good example is VIRGOHI21 (Davies et al. 2004, Minchin et al. 2005).
This very massive (from dynamical estimates $\approx 10^{11}$
M$_{\sun}$) dark galaxy could either be the result of a high
velocity encounter (Duc \& Bournaud 2008) or simply gas trapped in a
dark matter halo, not yet forming stars. Taking the results of this
simulation into account another possibility arise, VIRGOHI21 could
be the result of a strong stripping event. Like in Fig.
\ref{all_dmw_5_new} nearly to whole gaseous disc could be stripped
by a single stripping event. The gas complex, as not bound to a dark
matter halo, would not form stars on time scales comparable to the
same amount of gas trapped in a dark matter halo.

\section{Discussion and Conclusions}
We present numerical studies of the influence of ram pressure on
disc galaxies and their stellar and gaseous components. We performed
12 different high resolution SPH simulations with the same mass
resolution for the gas in the disc and the surrounding gas. The
simulations include radiative cooling and a recipe for star
formation (Springel \& Hernquist, 2003). We varied the density of
the surrounding gas from $1\times10^{-28}$ to $5\times10^{-27}$
g/cm$^{3}$ and the relative velocity of the galaxy with respect to
the surrounding gas from 100 km/s to 1000 km/s. The chosen
combinations probe the ram pressure range from $1\times10^{-12}$ to
$5\times10^{-13}$ dyn cm$^2$. The surrounding gas temperature was
chosen to be $1\times10^7$ K, this leads to ICM conditions similar
to many regions in the Virgo galaxy cluster (Schindler et al. 1999).
The study here can be seen as an extension of the studies presented
by Kronberger et al. (2008) and Kapferer et al. (2008), including
mock observations in different wavelength, like in the HI, H$\alpha$
and X-ray range. The results of this work can be summarised as
follows:
\begin{itemize}

\item The  star-formation rate for simulations with ram
pressure can be more then 10 times higher then for the same galaxy
evolving in vacuum. The star-formation rate depends more strongly on
the surrounding gas density than on the relative velocity. This can
be explained by the location of the newly formed stars. The higher
the ram pressure, the more gas is stripped from the galaxy, leading
to less star formation in the disc. On the other hand more
star-forming gaseous-knots are present in the gaseous wake, which
form many new stars. The higher the external pressure on this knots,
the more stars are formed. For surrounding gas densities above
$5\times10^{-27}$ g/cm$^{3}$ the star formation is the highest for
the lowest relative velocity. For lower gas densities this trend reverses.\\

\item The distribution of newly formed stars depends strongly on the
ram pressure on the galaxy. More than 95\% of all the newly formed
stars in the simulation with the maximum ram pressure are located in
the stripped wake of the galaxy (simulation 4). The stars are formed
from the stripped gas originating from the galaxy at larger
distances, up to 400 kpc behind the disc after 500 Myr. If the
relative velocity drops below 500 km/s and the surrounding gas
density is below $\rho \leq 1\times10^{-28}$ g/cm$^{3}$ all the new
stars are formed in the disc.\\

\item The surface density of stars in the wake is typically 3 to 4
magnitudes lower than in the central regions of the stellar disc.
The locations of newly formed stars can be found hundreds of kpc
behind the disc, depending on the strength and duration of the ram
pressure. Due to the fact that newly formed stars are not feeling
the ram pressure anymore, their trajectories are only influenced by
the underlying gravitational potential. They fall back to the centre
of the mass distribution, which is the stellar disc. As the stars
are treated as a collisionless fluid they are falling through the
stellar disc and appear in the opposite direction of the wake. This
leads to stellar bulges around the old stellar disc, they are more
pronounced in the opposite
direction of the ram pressure.\\

\item A young stellar population at a given time (formed within
the last 50 Myr), which could be called OB-associations, is
detectable throughout the whole wake. In the initial ram pressure
event, the first 50 to 200 Myr depending on the strength of the ram
pressure, they are very prominent in the whole stellar component
present in the wake. After the initial stripping event, the
OB-associations are found at the locations of the largest gaseous
densities in the wake, the regions were the new stars are formed.
Their surface density is locally of the same order of magnitude as
in the stellar disc, if the ram pressure does not strip the whole
gas of the disc.\\

\item If the model galaxy is affected by strong external ram pressure the
positions of star formation are shifted from the disc to the wake.
In observations this would lead to a classification of the disc
galaxy as a passively evolving system, although the overall star
formation rate (disc and wake) is strongly enhanced by the ram
pressure. The time scale can be very short, in our strongest
ram-pressure stripping simulation it is less than 80 Myr.\\

\item The influence of ram pressure on the gaseous component
depends strongly on the strength of the ram pressure. In the case of
simulation 4 (v$_{\rm{rel}}=1000$ km/s and a surrounding gas density
of $5\times10^{-27}$ g/cm$^{3}$) the gaseous disc is stripped
completely after 75 Myr of ram pressure. By reducing the relative
velocity by a factor of 2 (500 km/s) the disc is stripped off more
then 80\% of the initial gas mass in the disc. As the gas in the
galaxy has an internal velocity pattern from the rotational velocity
field of the galaxy, the stripped gas has complex structures in the
stripped wake. As a result, the stripped wake forms filaments and
gas concentrations. The self-gravitating gas concentrations survive
throughout the whole simulation time (more than 750 Myr).\\

\item The number of dense gaseous knots, which act as star-forming
cocoons increases with the strength of the ram pressure. In the
simulation with the highest ram pressure (v$_{\rm{rel}}$=1000 km/s
$\rho_{\rm{sur}}=1\times10^{-27}$ g/cm$^{3}$) the number of gaseous
knots with a surface density above $1\times10^5$ M$_{\sun}$/kpc$^3$
is three times higher than in the case with
$\rho_{\rm{sur}}=1\times10^{-28}$ g/cm$^{3}$.\\

\item The mock X-ray observations of the simulations reveal X-ray
tails with lengths of several hundred kpc, depending on the strength
of the ram pressure. Complex structures can be found in the X-ray
surface brightness, temperature and metal maps. One common feature
is a bright spot in all X-ray maps behind the stellar disc at
distances of a few tens of kpc caused by compressional heating of
the stripped gas. The location of the initial stripping event can be
seen nicely as the second bright feature in the X-ray maps. The
stronger the ram pressure, the more pronounced the feature. In the
simulations with 1000 km/s and a surrounding gas density of
$1\times10^{-28}$ g/cm$^{3}$ the X-ray tail is longer than in the
case with ten times higher surrounding gas density. This leads to
the conclusion that a strong ram-pressure event, stripping off the
whole gaseous disc within the initial stripping event, produces
metal structures in the X-ray maps which are spatially more
concentrated than in events with less ram pressure.\\

\item The cold stripped gas is not as widely spread in the wake as the
X-ray emitting hot gas. It can be mainly found near the location of
the first stripping event and near the stripped disc. Even features
in the opposite direction of the wake can be found. The reaccretion
of cold gas in the wake, which falls back onto the disc as it enters
the slipstream of the remaining gas in the disc is the explanation
for this. If the gaseous disc is completely stripped, cold gas is
spread over more then 150 kpc around the maxima in the X-ray surface
brightness, temperature and metal map.\\
\end{itemize}
Ram pressure stripping changes the stellar- and the gaseous
distribution in galaxies drastically. Therefore the investigation of
this external process is crucial for understanding galaxy evolution
in galaxy clusters or groups. As the timescales occurring in
ram-pressure stripping events are fast compared to internal
processes in the disc, the transition from an actively to a
passively evolving system can occur on very short timescales. Due to
observational limits the conclusion that a system is passively
evolving can be misleading as the total star formation is typically
enhanced in a ram-pressure event, but the location of star formation
is moved from the disc into the stripped wake. Another implication
of the presented results affects observations of metal enrichment
processes in the ICM. The element abundances originating from
galactic winds and ram pressure stripping are maybe very similar.
The presence of ram-pressure induced intra-cluster star formation
makes it very difficult to distinguish between gas removed by
galactic winds or gas removed by ram pressure. Further
investigations are needed to disentangle the influence of the
different processes on the
chemical composition of the intra-cluster medium.\\
How other effects, like magnetic fields or different heating and
cooling processes, due to a complex mixture of different gas phases,
would change the presented numbers needs further investigations as
well. Recent observational findings and the presented simulations
here reveal a luminous universe outside the bright spots, the
galaxies.

\section*{Acknowledgements}
The authors would like to thank the referee Curtis Struck for his
constructive suggestions, which increased the quality of the paper.
The authors thank Volker Springel for providing them with {\sc
GADGET2}. The authors are grateful to Marco Barden for many useful
discussions and for Sabine Kreidl and Gerhard Niederwieser for the
perfect support at the Computational Centre at the University of
Innsbruck. The authors acknowledge the Austrian Science Foundation
(FWF) through grants P18523-N16 and P19300-N16. The authors further
acknowledge the UniInfrastrukturprogramm des BMWF Forschungsprojekt
Konsortium Hochleistungsrechnen.


\begin{thebibliography}{} 
\bibliographystyle{aa}

\bibitem[Abadi et al.(1999)]{1999MNRAS.308..947A} Abadi, M.~G., Moore, B.,
\& Bower, R.~G.\ 1999, \mnras, 308, 947

\bibitem[Arnaboldi et al.(2003)]{2003AJ....125..514A} Arnaboldi, M., et
al.\ 2003, \aj, 125, 514

\bibitem[Balogh et al.(2002)]{2002MNRAS.337..256B} Balogh, M., Bower,
R.~G., Smail, I., Ziegler, B.~L., Davies, R.~L., Gaztelu, A., \&
Fritz, A.\ 2002, \mnras, 337, 256

\bibitem[1991]{Breitschwerdt}
Breitschwerdt, D., McKenzie, J.F., V\"olk, H.J., 1991, A\&A 245, 79

\bibitem[Butcher
\& Oemler(1978)]{1978ApJ...219...18B} Butcher, H., \& Oemler, A.,
Jr.\ 1978, \apj, 219, 18

\bibitem[Chemin et al.(2006)]{2006MNRAS.366..812C} Chemin, L., et al.\
2006, \mnras, 366, 812

\bibitem[Davies et al.(2004)]{2004MNRAS.349..922D} Davies, J., et al.\
2004, \mnras, 349, 922

\bibitem[Domainko et al.(2006)]{2006A&A...452..795D} Domainko, W., et al.\ 2006, \aap,
452, 795

\bibitem[Dressler(1987)]{1987nngp.proc..276D} Dressler, A.\ 1987,
Nearly Normal Galaxies: From the Planck Time to the Present, ed. S.
M. Faber (New York: Springer-Verlag), p. 163.276

\bibitem[Dressler et al.(1999)]{1999ApJS..122...51D} Dressler, A., Smail,
I., Poggianti, B.~M., Butcher, H., Couch, W.~J., Ellis, R.~S., \&
Oemler, A.~J.\ 1999, \apjs, 122, 51

\bibitem[Duc \& Bournaud(2008)]{2008ApJ...673..787D} Duc, P.-A., \& Bournaud, F.\
2008, \apj, 673, 787

\bibitem[Feldmeier et al.(1998)]{1998ApJ...503..109F}
Feldmeier, J.~J., Ciardullo, R., \& Jacoby, G.~H.\ 1998, \apj, 503,
109

\bibitem[Ganda et al.(2006)]{2006MNRAS.367...46G} Ganda, K.,
Falc{\'o}n-Barroso, J., Peletier, R.~F., Cappellari, M., Emsellem,
E., McDermid, R.~M., de Zeeuw, P.~T., \& Carollo, C.~M.\ 2006,
\mnras, 367, 46

\bibitem[Garrido et
al.(2002)]{2002A&A...387..821G} Garrido, O., Marcelin, M., Amram,
P., \& Boulesteix, J.\ 2002, \aap, 387, 821

\bibitem[Gerken et al.(2004)]{2004A&A...421...59G} Gerken, B., Ziegler, B., Balogh, M.,
Gilbank, D., Fritz, A., J\"ager, K.\ 2004, \aap, 421, 59

\bibitem[Gingold \& Monaghan(1977)]{1977MNRAS.181..375G} Gingold,
R.~A., \& Monaghan, J.~J.\ 1977, \mnras, 181, 375

\bibitem[Gunn \& Gott(1972)]{1972ApJ...176....1G} Gunn, J.~E., \& Gott, J.~R.~I.\
1972, \apj, 176, 1

\bibitem[J{\'a}chym et
al.(2007)]{2007A&A...472....5J} J{\'a}chym, P., Palou{\v s}, J.,
K{\"o}ppen, J., \& Combes, F.\ 2007, \aap, 472, 5

\bibitem[Kapferer et al.(2006)]{2006A&A...446..847K} Kapferer, W., Kronberger, T.,
Schindler, S., B{\"o}hm, A., \& Ziegler, B.~L.\ 2006, \aap, 446, 847

\bibitem[Kapferer et
al.(2007)]{2007A&A...466..813K} Kapferer, W., et al.\ 2007, \aap,
466, 813

\bibitem[Kapferer et
al.(2007)]{2007A&A...472..757K} Kapferer, W., Kronberger, T.,
Weratschnig, J., \& Schindler, S.\ 2007, \aap, 472, 757

\bibitem[Kapferer et al.(2008)]{2008MNRAS.389.1405K} Kapferer, W.,
Kronberger, T., Ferrari, C., Riser, T.,
\& Schindler, S.\ 2008, \mnras, 389, 1405

\bibitem[Kim et al.(2008)]{2008ApJ...688..931K} Kim, D.-W., Kim, E.,
Fabbiano, G., \& Trinchieri, G.\ 2008, \apj, 688, 931

\bibitem[Kronberger et
al.(2006)]{2006A&A...458...69K} Kronberger, T., Kapferer, W.,
Schindler, S., B{\"o}hm, A., Kutdemir, E., \& Ziegler, B.~L.\ 2006,
\aap, 458, 69

\bibitem[Kronberger et
al.(2007)]{2007A&A...473..761K} Kronberger, T., Kapferer, W.,
Schindler, S., \& Ziegler, B.~L.\ 2007, \aap, 473, 761

\bibitem[Kronberger et al.(2008)]{2008A&A...481..337K} Kronberger, T., Kapferer, W., Ferrari, C., Unterguggenberger, S., \&
    Schindler, S.\ 2008, \aap, 481, 337

\bibitem[Larson et al.(1980)]{1980ApJ...237..692L} Larson, R.~B., Tinsley,
B.~M., \& Caldwell, C.~N.\ 1980, \apj, 237, 692

\bibitem[Lucy(1977)]{1977AJ.....82.1013L} Lucy, L.~B.\ 1977, \aj, 82, 1013

\bibitem[Maurogordato et al.(2008)]{2008A&A...481..593M} Maurogordato, S., et al.\ 2008, \aap, 481, 593

\bibitem[Mihos et al.(2005)]{2005AAS...20717702M} Mihos, C., Harding, P.,
Feldmeier, J., \& Morrison, H.\ 2005, Bulletin of the American
Astronomical Society, 37, 1449

\bibitem[Minchin et al.(2005)]{2005ApJ...622L..21M} Minchin, R., et al.\
2005, \apjl, 622, L21

\bibitem[Moore et al.(1998)]{1998ApJ...495..139M} Moore, B., Lake, G.,
\& Katz, N.\ 1998, \apj, 495, 139

\bibitem[Mori
\& Burkert(2000)]{2000ApJ...538..559M} Mori, M., \& Burkert, A.\
2000, \apj, 538, 559

\bibitem[Poggianti et al.(2006)]{2006ApJ...642..188P} Poggianti, B.~M., et
al.\ 2006, \apj, 642, 188

\bibitem[Quilis et al.(2000)]{2000Sci...288.1617Q} Quilis, V., Moore, B.,
\& Bower, R.\ 2000, Science, 288, 1617

\bibitem[Randall et al.(2008)]{2008ApJ...688..208R} Randall, S., Nulsen,
P., Forman, W.~R., Jones, C., Machacek, M., Murray, S.~S.,
\& Maughan, B.\ 2008, \apj, 688, 208

\bibitem[Roediger
\& Hensler(2005)]{2005A&A...433..875R} Roediger, E., \& Hensler, G.\
2005, \aap, 433, 875

\bibitem[Roediger
\& Br{\"u}ggen(2006)]{2006MNRAS.369..567R} Roediger, E., \&
Br{\"u}ggen, M.\ 2006, \mnras, 369, 567

\bibitem[Roediger
\& Br{\"u}ggen(2007)]{2007MNRAS.380.1399R} Roediger, E., \&
Br{\"u}ggen, M.\ 2007, \mnras, 380, 1399

\bibitem[Roediger
\& Br{\"u}ggen(2008)]{2008MNRAS.388..465R} Roediger, E., \&
Br{\"u}ggen, M.\ 2008, \mnras, 388, 465

\bibitem[Rubin et al.(1999)]{1999AJ....118..236R} Rubin, V.~C., Waterman,
A.~H., \& Kenney, J.~D.~P.\ 1999, \aj, 118, 236

\bibitem[Russell et al.(2008)]{2008MNRAS.390.1207R} Russell, H.~R.,
Sanders, J.~S., \& Fabian, A.~C.\ 2008, \mnras, 390, 1207

\bibitem[Schindler et al.(1999)]{1999A&A...343..420S} Schindler, S., Binggeli, B., B\"ohringer, H.\ 1999, \aap, 343, 420

\bibitem[Schindler et
al.(2005)]{2005A&A...435L..25S} Schindler, S., et al.\ 2005, \aap,
435, L25

\bibitem[Schulz \& Struck(2001)]{2001MNRAS.328..185S} Schulz, S., \& Struck, C.\
2001, \mnras, 328, 185

\bibitem[Springel \& Hernquist(2003)]{2003MNRAS.339..289S} Springel, V., \&
Hernquist, L.\ 2003, \mnras, 339, 289

\bibitem[Springel(2005)]{2005MNRAS.364.1105S} Springel, V.\ 2005, \mnras,
364, 1105

\bibitem[Sun et al.(2007)]{2007ApJ...671..190S} Sun, M., Donahue, M.,
\& Voit, G.~M.\ 2007, \apj, 671, 190

\bibitem[Theuns
\& Warren(1997)]{1997MNRAS.284L..11T} Theuns, T., \& Warren, S.~J.\
1997, \mnras, 284, L11

\bibitem[Toniazzo
\& Schindler(2001)]{2001MNRAS.325..509T} Toniazzo, T., \& Schindler,
S.\ 2001, \mnras, 325, 509

\bibitem[Verdugo et
al.(2008)]{2008A&A...486....9V} Verdugo, M., Ziegler, B.~L., \&
Gerken, B.\ 2008, \aap, 486, 9

\bibitem[Vollmer et al.(2001)]{2001ApJ...561..708V} Vollmer, B., Cayatte,
V., Balkowski, C., \& Duschl, W.~J.\ 2001, \apj, 561, 708

\bibitem[Yoshida et al.(2008)]{2008ApJ...688..918Y} Yoshida, M., et al.\
2008, \apj, 688, 918

\bibitem[Yoshida et al.(2004)]{2004AJ....127.3653Y} Yoshida, M., et al.\
2004, \aj, 127, 3653

\bibitem[Yoshida et al.(2002)]{2002ApJ...567..118Y} Yoshida, M., et al.\
2002, \apj, 567, 118


\end{thebibliography}
\end{document}